\documentclass[journal]{IEEEtran}
\usepackage{latexsym}

\usepackage{enumerate}
\usepackage{graphicx}
\usepackage{subfigure}
\usepackage{multirow}

\usepackage{framed,multirow}
\usepackage{times}
\usepackage{graphicx} % more modern
\usepackage{subfigure}
\usepackage{amssymb}
\usepackage{latexsym}

\usepackage{algorithm}
\usepackage{algorithmic}

\usepackage{hyperref}

\usepackage{url}
\usepackage{xcolor}
\usepackage{amsmath}
\usepackage{amsxtra}
\usepackage{braket}
\usepackage{array}
\usepackage{color}
\usepackage{url}
\usepackage{placeins}

\usepackage{times}
\usepackage{balance}
\usepackage{enumerate}
\usepackage{multirow}
\usepackage{pifont}
\usepackage{marvosym}
\usepackage{ifsym}

\begin{document}
\title{A Quantum-inspired Similarity Measure for the Analysis of Complete Weighted Graphs}
\author{Lu~Bai,~\IEEEmembership{}Luca~Rossi,~\IEEEmembership{}Lixin~Cui,~\IEEEmembership{}Jian Cheng,~\IEEEmembership{}Edwin~R.~Hancock,~\IEEEmembership{IEEE~Fellow}

\thanks{Lu Bai is with Central University of Finance and Economics, Beijing, China. e-mail: bailucs@cufe.edu.cn; bailu69@hotmail.com.}
\thanks{Luca Rossi is with Southern University of Science and Technology, Guangdong, China. email: rossil@sustech.edu.cn (Co-Corresponding Author)}
\thanks{Lixin Cui is with Central University of Finance and Economics, Beijing, China. e-mail: cuilixin@cufe.edu.cn (Corresponding Author)}
\thanks{Jian Cheng is with National Laboratory of Pattern Recognition, Institute of Automation, Chinese Academy of Sciences, Beijing, China.}
\thanks{Edwin R. Hancock is with University of York, York, UK. }}

\markboth{Journal of \LaTeX\ Class Files,~Vol.~6, No.~1, January~2007}%
{Shell \MakeLowercase{\textit{et al.}}: Bare Demo of IEEEtran.cls
for Journals}
\maketitle

\begin{abstract}
We develop a novel method for measuring the similarity between complete weighted graphs, which are probed by means of discrete-time quantum walks. Directly probing complete graphs using discrete-time quantum walks is intractable due to the cost of simulating the quantum walk. We overcome this problem by extracting a commute-time minimum spanning tree from the complete weighted graph. The spanning tree is probed by a discrete time quantum walk which is initialised using a weighted version of the Perron-Frobenius operator. This naturally encapsulates the edge weight information for the spanning tree extracted from the original graph. For each pair of complete weighted graphs to be compared, we simulate a discrete-time quantum walk on each of the corresponding commute time minimum spanning trees, and then compute the associated density matrices for the quantum walks. The probability of the walk visiting each edge of the spanning tree is given by the diagonal elements of the density matrices. The similarity between each pair of graphs is then computed using either a) the inner product or b) the negative exponential of the Jensen-Shannon divergence between the probability distributions. We show that in both cases the resulting similarity measure is positive definite and therefore corresponds to a kernel on the graphs. We perform a series of experiments on publicly available graph datasets from a variety of different domains, together with time-varying financial networks extracted from data for the New York Stock Exchange. Our experiments demonstrate the effectiveness of the proposed similarity measures.
\end{abstract}
% IEEEtran.cls defaults to using nonbold math in the Abstract.
% This preserves the distinction between vectors and scalars. However,
% if the journal you are submitting to favors bold math in the abstract,
% then you can use LaTeX's standard command \boldmath at the very start
% of the abstract to achieve this. Many IEEE journals frown on math
% in the abstract anyway.

% Note that keywords are not normally used for peerreview papers.

\IEEEpeerreviewmaketitle

\section{Introduction}
Graph-based representations commonly arise in a wide variety of systems that are naturally described in terms of relations between their components. For instance, Wu et al.~\cite{DBLP:journals/tcyb/WuPZC15} have represented the texts inside a webpage as graphs, with vertices representing words and edges denoting relations between words. Li et al.~\cite{DBLP:journals/tcyb/LiHWL16} have represented each video frame as a graph structure with vertices representing superpixels and edges denoting relations between superpixels. Tang et al.~\cite{DBLP:journals/tcyb/TangSLL16} have looked at the local spectral descriptors of each image as points and constructed the weighted graph based on neighborhood relations. Other typical examples include representing a chemical molecular structure or a document as graphs~\cite{DBLP:journals/tcyb/XuanLZL15}. One main problem of analyzing these structures is that of measuring the similarity between two graphs for classification or clustering~\cite{DBLP:journals/tsmc/RiesenB09,DBLP:journals/tsmc/SanfeliuF83}. For example, in network science a common objective is to detect the extreme events that can significantly change the time-varying network structures~\cite{NetworkScience1,NetworkScience2,NetworkScience3,DBLP:series/sci/BunkeDHIK07,DBLP:journals/join/ShoubridgeKWB02,DBLP:conf/ibpria/BunkeDK05,DBLP:conf/mldm/BunkeDIK05} abstracted from vectorial time series~\cite{bullmore2009complex}. Given a measure of similarity between graphs, one can then identify extreme events by looking at significant changes in the underlying network structures.

Consider a system that can be represented by a series of complete weighted graphs, where the number of vertices $N$ is fixed but the edge weights change over time. Such a representation is a natural choice for those systems where we are interested in modelling the strength of the edges rather than their presence or absence. With this time-varying network to hand, one can look at how it evolves in order to detect anomalies in the system. An example of this type of graph is furnished by financial networks, where each vertex represents a stock and each edge weight measures the association between the time series of the corresponding stock, in terms of correlation~\cite{ye2015thermodynamic}, Granger causality~\cite{vyrost2015granger} or transfer entropy~\cite{sandoval2014structure}. In this domain, extreme events representing financial instability of different stock are of interest~\cite{silva2015modular} and can be inferred by detecting the anomalies in the corresponding networks~\cite{ye2015thermodynamic}.

Existing methods aim to derive network characteristics based on connectivity structure, or statistics capturing connectivity structure~\cite{feldman1998measures,anand2011shannon,anand2014entropy}. These methods focus on capturing network substructures using clusters, hubs and communities. Moreover, an alternative principled approach is to characterize the networks using ideas from statistical physics~\cite{BOOK1,javarone2013quantum}. These methods use the partition function to describe the network, and the associated entropy, energy, and temperature measures can be computed through this function~\cite{ye2015thermodynamic,delvenne2011centrality,fronczak2007thermodynamic}. Unfortunately, the aforementioned methods tend to characterize network structures in a low dimensional pattern or vector space, and thus discard a lot of structural information. This drawback influences the effectiveness of existing approaches for time-varying network analysis. The aim of this paper is to address this shortcoming of existing methods by means of graph kernels.

\subsection{Graph Kernels}

In machine learning, graph kernels are important tools for analyzing structured data that are represented by graphs. This is because graph kernels not only allow us to map graph structures into a high dimensional space, but also provide a way of making the rapidly developing kernel machinery for vectorial data applicable to graphs. In essence, graph kernels are positive definite similarity measures between pairs of graphs~\cite{DBLP:journals/tsmc/RiesenB09,DBLP:conf/icdm/BorgwardtK05,GarterCOLT2003,DBLP:conf/icml/KashimaTI03,DBLP:journals/pr/Bai16,DBLP:journals/jmlr/ShervashidzeVPMB09,DBLP:conf/icml/Bai0ZH15,DBLP:conf/cvpr/HarchaouiB07}. They allow rapidly developing kernelized algorithms (e.g., Support Vector Machines, kernel Principle Component Analysis, etc) for vectorial data associating with vectorial kernels~\cite{DBLP:journals/tcyb/HanYML14,DBLP:journals/tcyb/LiuWYZZ13} directly applicable to graph structures.

One leading principle for defining kernels between a pair of graphs is to decompose the graphs into substructures and to measure the similarity between the input graphs by enumerating pairs of isomorphic substructures. Specifically, any available graph decomposition method can be adapted to develop a graph kernel, e.g., graph kernels based on counting pairs of isomorphic a) paths~\cite{DBLP:conf/icdm/BorgwardtK05}, b) walks~\cite{GarterCOLT2003,DBLP:conf/icml/KashimaTI03},  and c) subgraphs or subtrees~\cite{DBLP:journals/pr/Bai16,DBLP:journals/jmlr/ShervashidzeVPMB09,DBLP:conf/icml/Bai0ZH15,DBLP:conf/cvpr/HarchaouiB07}. Unfortunately, there are two common shortcomings arising with these graph kernels. First, they do not work well on complete weighted graphs, where each pair of vertices is linked by a weighted edge. This is due to the trivial structure of the complete graph, i.e., each vertex is adjacent to all the other vertices, whereas the weights may be quite different. Thus, it is difficult to decompose a complete weighted graph into substructures. On the other hand, identifying the isomorphism between weighted (sub)graphs tends to be computationally burdensome unless the weight information is discarded. Second, graph kernels cannot scale up to large structures. To overcome this issue, existing graph kernels usually compromise and use small sized substructures. However, measuring kernel values with small substructures only partially reflects the topological characteristics of a graph.

To address the restriction of R-convolution graph kernels to complete weighted graphs, a number of graph kernels~\cite{DBLP:conf/icml/JohanssonJDB14,xu2015local,DBLP:journals/jmiv/BaiH13} based on using the adjacency matrix to capture global graph characteristics have been developed. Since the adjacency matrix directly reflects the edge weight information, these kernels can naturally accommodate complete weighted graphs. For instance, Johansson~\cite{DBLP:conf/icml/JohanssonJDB14} et al. have developed a family of global graph kernels based on the Lov¡äasz number and its associated orthonormal representation through the adjacency matrix. Xu et al.~\cite{xu2015local} have proposed a local-global mixed reproducing kernel based on the approximated von Neumann entropy through the adjacency matrix. Bai et al.~\cite{DBLP:journals/jmiv/BaiH13} have defined an information theoretic kernel based on the classical Jensen-Shannon divergence between the steady state random walk probability distributions obtained through the adjacency matrix. Recently, there has been increasing interests in continuous-time quantum walks~\cite{quantum1998} for the analysis of global graph structures. The continuous-time quantum walk is the quantum analogue of the classical continuous-time random walk. Unlike the classical random walk that is governed by a doubly stochastic matrix, the quantum walk is governed by a unitary matrix and is not dominated by the low frequencies of the Laplacian spectrum. Thus, the continuous-time quantum walk is able to better discriminate different graph structures.

There have been a number of graph kernels developed using the continuous-time quantum walk. For instance, Bai et al.~\cite{DBLP:journals/pr/Bai0TH15} have developed a quantum kernel by measuring the similarity between two continuous-time quantum walks
evolving on a pair of graphs. Specifically, they associate each graph with a mixed quantum state that represents the evolution of the quantum walk. The resulting kernel is computed by measuring the quantum Jensen-Shannon divergence between the associated density matrices. Rossi et al.~\cite{rossi2015measuring} have developed a quantum kernel by exploiting the
relation between the continuous-time quantum walk intereferences and the symmetries of a pair of graphs, in terms of the quantum Jensen-Shannon divergence. Both of these quantum kernels employ the Laplacian matrix as the required Hamiltonian operator, and thus they can naturally accommodate complete weighted graphs. Unfortunately, all of the aforementioned kernels, either the classical kernels defined through the adjacency matrix or the quantum kernels based on the Laplacian matrix as the Hamiltonian operator, are restricted to un-attributed graphs. Furthermore, both quantum kernels require a composite structure of each pair of graphs to compute an additional mixed state that describes a system having equal probability of being in the two original quantum states. Unless, we can make use of the transitive alignment information between the vertices of the two graphs, then neither of the quantum kernels are positive definite.

\subsection{Contributions}\label{s1.2}
\begin{figure*}[t!]
\vspace{-0pt}
\centering
\subfigure{\includegraphics[width=0.8\linewidth]{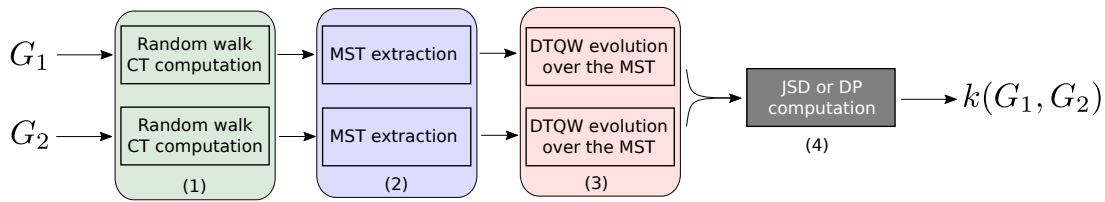}}
\vspace{-10pt}
\caption{The proposed framework to compute the similarity between two complete weighted graphs. Given two input graphs, for each of them (1) we compute the random walk commute time matrix, and (2) we extract the corresponding commute time spanning tree. (3) We probe the structure of each tree using discrete-time quantum walks and computing the time-average probability of visiting each arc residing on the edge. Finally, (4) the kernel between the original input graphs is defined as the similarity between the corresponding time-averaged probability distributions.}
\vspace{-10pt}
\label{pipeline}
\end{figure*}

In this paper, we aim to address the above mentioned drawbacks of existing state-of-the-art graph kernels by proposing a new kernel for complete weighted graphs. We test this similarity measure on graphs extracted from financial time series, as these are typically abstracted using sets of complete weighted graphs. We stress, however, that our kernel can also be applied to general graphs. In Section~\ref{s5} we measure the performance of the new kernel on a series of graph embedding and classification tasks, showing that it significantly outperforms a number of widely used alternative kernels.

Given a set of graphs, our aim is to probe each of them using a discrete-time quantum walk~\cite{DBLP:journals/pr/EmmsSWH09}. The reasons for using a walk of this type are twofold. First, quantum walks possess several exotic properties not exhibited by their classical counterpart. These in turn are a consequence of the marked differences between the two types of walks. For instance, unlike the classical walks the evolution of which is governed by a stochastic matrix, the behaviour of quantum walks is governed by a unitary matrix. As a result of their unique properties, during their evolution, quantum walks produce destructive interference patterns which have been shown to lead to more powerful graph characterisations~\cite{DBLP:journals/pr/EmmsSWH09}. Second, unlike continuous-time quantum walks~\cite{DBLP:conf/gbrpr/Bai0RZH15}, the state space of a discrete-time quantum walk is the set of directed edges rather than the set of vertices. In particular, suppose $G$ is a sample graph with vertex set $V$ and edge set $E$. We replace each edge $\{u,v\}\in E$ with a pair of directed edges $e_d(u,v)$ and $e_d(v,u)$, and we denote this set as $E_d$. Since $E_d$ is the state space of the discrete-time quantum walk and its cardinality is much larger than, or at least equal to, that of the vertex set $V$, the discrete-time quantum walk can capture the structural characteristics of the graph better than its continuous-time counter part.

Unfortunately, directly simulating the discrete-time quantum walk on a complete weighted graph tends to be a challenging problem. The time complexity of simulating the evolution of a discrete-time quantum walk is cubic in the size of the state space. For a complete weighted graph having $n$ vertices, there are ${n(n-1)}$ directed edges. Thus, performing the discrete-time quantum walk on a complete weighted graph has an associated time complexity of $O(n^6)$. Consequently, the use of discrete-time quantum walks for the analysis of complete weighted graphs is computationally expensive and unadvisable. Most importantly, two complete weighted graphs over the same set of vertices are clearly structurally equivalent. In other words, unless we consider the edge weighted information, these graphs are virtually indistinguishable.

One way to overcome these issues is to compute sparser versions of the original graphs. There have been a number of alternative approaches to extract sparse structures from complete or dense weighted graphs. For instance, Ye et al.~\cite{ye2015thermodynamic} and Silva et al.~\cite{silva2015modular} take the widely used threshold-based approach. That is, they preserve only those edges whose weights fall into the larger $10\%$ of weights. Unfortunately, these methods usually lead to a significant loss of information, since many weighted edges are discarded. Moreover, the resulting structure clearly depends on the choice of the threshold, and it is generally unclear how this threshold should be selected. This in turn results in very unstable and potentially disjoint graphs. Mantegna and Stanley~\cite{mantegna1999introduction} have extracted the minimum spanning tree over the original weighted adjacency matrix. This removes the need of selecting a threshold and the resulting instability, but it still suffers from the information loss problem of threshold-based methods.

To address the aforementioned problems, in this paper we propose to sparsify the original graph using the minimum spanning tree through the commute time matrix rather than the original weighted adjacency matrix. The commute time averages the time taken for a random walk to travel between a pair of vertices over all connecting paths, and is robust to the deletion of individual edges or paths (i.e., structural noise) unless these form bridges between connected components of a graph~\cite{DBLP:journals/pami/QiuH07a}. Thus, the resulting spanning tree structure through the commute time matrix not only retains salient structural characteristics of the graph, but also encapsulates proximity information residing on the discarded edges. In other words, we minimize the edge number in the original graph while preserving most of its path structure information. This, as shown in~\cite{DBLP:journals/prl/Bai16}, can lead to a significant reduction of the computational complexity when applied to dense graphs. Most importantly, the commute time can easily accommodate weighted information residing on edges. As a result, the commute time not only represents an ideal candidate for sparsifying the structure of the original graph~\cite{DBLP:journals/pr/QiuH07b}, but also allows us to separate otherwise structurally indistinguishable complete graphs.

The aim of this paper is to develop a new kernel for complete weighted graphs associated with discrete-time quantum walks. To this end, we propose a new framework of computing this kernel and proceed as follows. Given the commute time spanning tree representations of the original pair of graphs, we first simulate the evolution of a discrete-time quantum walk on each of the trees, where we make use of a novel weighted version of the Perron-Frobenius operator~\cite{PengQuantum2011}. This in turn allows us to encode the weights on the edges of the commute time spanning tree in the initial state of the walk. Then, for each discrete-time quantum walk, we compute the associated time-averaged density matrix. Density matrices are matrices that describe quantum systems that are in a statistical mixture of quantum states, and they play a fundamental role in the quantum observation process. In our case, the time-averaged density matrix describes a statistical ensemble of quantum states encapsulating the time-evolution of a quantum walk. The diagonal of this matrix corresponds to the time-averaged probability distribution of the walk visiting each arc residing on the edge of the underlying graph. With a pair of density matrices to hand, the kernel between the original graphs is computed as the similarity between the associated time-averaged probability distributions. We show that the similarity between these distributions can be computed either as the negative exponential of their classical Jensen-Shannon divergence~\cite{majtey2005jensen} or as their dot product. Both approaches lead to the definition of a positive definite kernel measure. Fig.\ref{pipeline} illustrates the structure of the proposed framework to compute the kernel based similarity between two complete weighted graphs. Experiments on financial networks datasets as well as standard graph datasets abstracted from the bioinformatics domain demonstrate the effectiveness of the new kernel.

Finally, note that the proposed kernel is closely related to the kernel we introduced in~\cite{DBLP:journals/prl/Bai16}, where we proposed to simplify the structure of the input graphs through commute time, to then compare them using discrete-time quantum walks. However, the proposed kernel significantly differs from~\cite{DBLP:journals/prl/Bai16} and has a number of important theoretical advantages. First, the computation of the initial quantum state for the proposed kernel is based on the newly introduced weighted Perron-Frobenius operator. As a result, only the proposed kernel can encapsulate the edge weight information of the original graphs. Second, unlike~\cite{DBLP:journals/prl/Bai16}, where the similarity between the input graphs is computed using the quantum Jensen-Shannon divergence between the density matrices associated with the graphs, here we look at the classical Jensen-Shannon divergence between the probability distributions associated with these density matrices. In particular, we show that these probability distributions can be easily transformed to distributions over the space of directed edge labels, defined as the union of the original edge and vertex labels, thus allowing this new kernel to incorporate both vertex and edge labels of the original graphs. Finally, in order to compute the divergence between the full density matrices, the kernel in~\cite{DBLP:journals/prl/Bai16} also requires the computation of an additional mixed density matrix for each pair of input graphs. This does not take into account the correspondences between the nodes of the input graphs and thus does not guarantee permutation invariance. On the other hand, we overcome this problem by computing the divergence between probability distributions over a common state space, i.e., the space of directed edge labels. As a result, the proposed kernel is both permutation invariant and positive definite.

The remainder of the paper is organised as follows. Section~\ref{s2} introduces the necessary quantum mechanical background, while Section~\ref{s3} reviews the concept of commute time and shows how to sparsify a graph using the commute time spanning tree. Section~\ref{s4} introduces the proposed kernel, which is extensively evaluated in Section~\ref{s5}. Finally, Section~\ref{s6} concludes the paper.

\section{Quantum Mechanical Background}\label{s2}

\subsection{Discrete-time Quantum Walks}
In quantum mechanics, discrete-time quantum walks are defined as the quantum counterparts of classical discrete-time random walks~\cite{DBLP:journals/pr/EmmsSWH09}. Quantum processes are reversible, so in quantum walks the states need to specify both the current and the previous locations of the walk. Let us replace each edge $e(u,v) \in E$ with a pair of directed edges $e_d(u,v)$ and $e_d(v,u)$, and denote the new set as $E_d$. The state space of the discrete-time quantum walk is $E_d$ and we represent the state of the quantum walker at $e_d(u,v)$ as $|uv\rangle$. That is, $|uv\rangle$ denotes the state in which the walk is at vertex $v$ having previously been at vertex $u$. A general state of the walk is
\begin{equation}
|\psi\rangle=\sum_{e_d(u,v)\in E_d}\alpha_{uv}|uv\rangle
\label{QuantumState}.
\end{equation}
\noindent
where the quantum amplitudes $\alpha_{uv}$ are complex. The probability that the walk is in state $|uv\rangle$ is given by $\mathrm{Pr}(|uv\rangle)=\alpha_{uv}\alpha_{uv}^*$, where $\alpha_{uv}^{*}$ is the complex conjugate of $\alpha_{uv}$.

At each time step, the quantum walk evolution is governed by the transition matrix $\textbf{\emph{U}}$, the entries of which indicate the transition probabilities between states, i.e.,
\begin{equation}
|\psi_{t+1}\rangle = \textbf{\emph{U}}|\psi_t\rangle.
\end{equation}
Since the walk evolution is linear and conserves probability, $\textbf{\emph{U}}$ must be an unitary matrix, i.e., the inverse of $\textbf{\emph{U}}$ is equal to its Hermitian transpose $\textbf{\emph{U}}^{\dagger}$. A typical choice is to choose the Grover matrix~\cite{Grover96} as the transition matrix, i.e.,
\begin{equation}
U_{(u,v),(w,x)}=\left\{
\begin{array}{ll}
\frac{2}{d_x}-\delta_{ux}, & \textrm{v=w;} \\
0,                         & \textrm{otherwise,}
\end{array}\right.
\label{EvolutionQuantumWalk}
\end{equation}
\noindent
where $d_x$ corresponds to the vertex degree of vertex $x$, $U_{(u,v),(w,x)}$ indicates the quantum amplitude of the transition $e_d(u,v) \rightarrow e_d(w,x)$, and $\delta_{ux}$ is the Kronecker delta, i.e., $\delta_{ux}=1$ if $u=x$ and $0$ otherwise. For each state $|u_1v\rangle$, $\textbf{\emph{U}}$ assigns the same amplitude to all transitions $|u_1v\rangle \rightarrow |vu_i\rangle$, and a different amplitude to the transition $|u_1v\rangle \rightarrow |vu_1\rangle$, where $u_i$ is a neighbour vertex of $v$. Note that the elements of $\textbf{\emph{U}}$ are real numbers, and they can be either positive or negative. This indicates that Eq.(\ref{EvolutionQuantumWalk}) allows \emph{destructive interference} to take place as a consequence of negative quantum amplitudes appearing during the walk evolution.

\subsection{The Weighted Perron-Frobenius Operator}

In~\cite{PengQuantum2011}, it was shown that there exist an important link between the Perron-Frobenius operator and the transition matrix of discrete-time quantum walks. To illustrate this linkage, we commence by introducing the concepts of directed line graph and positive support for a matrix.

\noindent\textbf{Definition 1} Let $G(V,E)$ be a given graph, the directed line graph $G_D(V_D,E_{D})$ of $G$ is a dual representation. To obtain $G_D$, we replace each edge $\{u,w\} \in E$ with a pair of directed edges $e_d(u,w)$ and $e_d(w,u)$ for vertices $u,w \in V$, and we denote this set as $E_d$. The directed line graph $G_D(V_D,E_{D})$ is an oriented graph with vertex set $V_D$ and edge set $E_{D}$ as
\begin{equation}
\left\{
\begin{array}{ll}
V_D =E_d, \\
E_{D}=\{(e_{d}(u,v), e_{d}(v,w)) \in E_d\times E_d\}.
\end{array}\right.\label{DOLG}
\end{equation}
where vertices $v,u,w \in V$ and $w \neq u$. Based on~\cite{PengQuantum2011}, the adjacency matrix $\textbf{\emph{T}}=[T_{i,j}]_{|V_D|\times |V_D|}$ of $G_D(V_D,E_{D})$ is the Perron-Frobenius operator.  \hfill$\square$

\noindent\textbf{Definition 2} Assume $\textbf{\emph{M}}=[M_{x,y}]_{m\times n}$ is a ${m\times n}$ matrix. Its positive support is a ${m\times n}$ matrix $\mathrm{S}^{+}(\textbf{\emph{M}})=[s_{x,y}]_{m\times n}$ with
\begin{equation}
s_{x,y}=\left\{
\begin{array}{l}
1  , \ \ \ \ \ \  M_{x,y}>0,\\
0  , \ \ \ \ \ \  \mathrm{otherwise},
\end{array}\right.
\end{equation}
where $1 \leq x \leq m$ and $1 \leq y \leq n$.\hfill$\square$

If $\textbf{\emph{U}}$ is the unitary matrix of a discrete-time quantum walk on the graph $G(V,E)$, then, based on~\cite{PengQuantum2011}, the Perron-Frobenius operator $\textbf{\emph{T}}$ of $G(V,E)$ can be constructed from the positive support of $\textbf{\emph{U}}$, i.e.,
\begin{equation}
\textbf{\emph{T}}=\textbf{\emph{S}}^{+}(\textbf{\emph{U}}^\top).\label{PFT}
\end{equation}
For the directed line graph $G_D(V_D,E_{D})$, each vertex $v_d\in V_D$ corresponds to a unique directed edge, thus the vertex set $V_D$ of $G_D$ essentially corresponds to the state space of the quantum walk. Furthermore, if there exists a directed edge from vertex $v_{d}\in V_D$ to vertex $u_{d}\in V_D$, the quantum walk on $G$ allows a transition from the directed edge representing $v_{d}$ to the directed edge representing $u_{d}$, and vice versa. These observations indicate that the discrete-time quantum walk on the original graph $G$ can be seen as a walk evolved on the corresponding directed line graph $G_D$, where the transitions of the walk are constrained by the directed edges of $G_D$.

The directed line graph has a number of interesting properties which in turn highlight some advantages of discrete-time quantum walks over their classical counterparts. First, the directed line graph can represent the original graph in a higher dimensional feature space. This is because the vertex set of the line graph corresponds to the set of directed edges of the original graph and its cardinality is usually greater than that of the vertex number of the original graph. Thus, compared to the continuous-time quantum walk~\cite{DBLP:journals/pr/Bai0TH15} on the original graph, the discrete-time quantum walk on the directed line graph can capture richer structural characteristics. Second, the directed line graph is a backtrackless representation of the original graph structure, because the edges of the line graph are all directed. Since the transitions of the discrete-time quantum walk are constrained by the directed edges of the line graph, the walk cannot visit a vertex and then immediately return to the starting vertex through the same edge. As a result, the discrete-time quantum walk can significantly reduce the notorious tottering problem of the classical random walk~\cite{DBLP:conf/icml/KashimaTI03}. Finally, since the discrete-time quantum walk and the directed line graph are related, the initial state of the quantum walk can be computed through the Perron-Frobenius operator $\emph{T}$ of the line graph~\cite{DBLP:journals/prl/Bai16}. Unfortunately, the operator $\emph{T}$ cannot reflect weight information residing on the edges of the original graph and thus its use leads to information loss. To overcome this shortcoming, we propose a new weighted Perron-Frobenius operator for the directed line graph. We compute the initial state by taking the square root of the sum of the out-degree and in-degree distributions from the new weighted operator.

\noindent\textbf{Definition 3 (Initial State from Directed Line Graphs)} Let $G(V,E)$ be a graph with weighted adjacency matrix $A$, $E_d$ the set of directed edges for $G$, and $G_D(V_D,E_{D})$ (\textbf{$E_d=V_D$}) the directed line graph of $G$. Each element of the weighted Perron-Frobenius operator $\textbf{\emph{T}}^W$ of $G$ satisfies
\begin{equation}
\textbf{\emph{T}}^W_{(u,v),(w,x)}=\left\{
\begin{array}{ll}
A_d(u,v)+A_d(w,x),  & {v=w}; \\
0,              & \textrm{otherwise,}
\end{array}\right.
\end{equation}
where $u,v,x, w\in V$, $A_d(u,v)=A(u,v)$ and $A_d(w,x)=A(w,x)$ are the weights of $e_d(u,v)\in E_d$ and $e_d(v,w)\in E_d$, $E_d= V_{D}$, and $(e_d(u,v),e_d(w,x))\in E_{D}$. %If $G$ is unweighted, $\textbf{\emph{T}}^W=\textbf{\emph{T}}$.
The initial state $\Ket{\psi_0}$ for $G$ through $G_D$ is
\begin{equation}
\Ket{\psi_0}=\sum_{e_d(u,v)\in E_d}\alpha_{uv}^0|uv\rangle
\label{InitialState},
\end{equation}
where
\begin{equation}
\alpha_{uv}^0= \sqrt{\frac{\sum_{e_d(w,x)\in V_L}\{\textbf{\emph{T}}^W_{(u,v),(w,x)}+\textbf{\emph{T}}^W_{(w,x),(u,v)}\}}
{\sum_{e_d(w,x),e_d(u,v)\in V_L}\{\textbf{\emph{T}}^W_{(u,v),(w,x)}+\textbf{\emph{T}}^W_{(w,x),(u,v)}\}}   }           .\label{IEvolutionQuantumWalk}
\end{equation}
The initial state $\Ket{\psi_0}$ not only preserves the structural information of $G_D$, but also encapsulates the edge weight information in the original graph $G$. \hfill$\square$

\subsection{From Quantum Walks to Density Matrices}\label{WPFO}
In quantum mechanics~\cite{PengQuantum2011}, a quantum system can be in a statistical ensemble of pure quantum states, where each pure state is described as a single ket vector $\Ket{\psi_i}$ and has an associated probability $p_i$. The density matrix of this quantum system is $\rho = \sum_i p_i \Ket{\psi_i}\Bra{\psi_i}$. Consider a sample graph $G(V,E)$ and let $\Ket{\psi_t}$ be the pure state that corresponds to a discrete-time quantum walk evolved from time step $t = 0$ to time step $t = T$ on $G$. The time-averaged density matrix $\rho_G^T$ for $G$ associated with the quantum walk is defined as
$\rho_G^T = \frac{1}{T+1} \sum_{t=0}^T \Ket{\psi_t}\Bra{\psi_t}$, since the state $\Ket{\psi_{t}}$ at time $t$ can be computed by $\Ket{\psi_{t}}=\textbf{\emph{U}}^t\Ket{\psi_{0}}$, where $\Ket{\psi_{0}}$ is the initial state of the quantum walk, and $\textbf{\emph{U}}$ is the transition matrix. Given the initial state $\Ket{\psi_0}$, $\rho_G^T$ can be re-written as
\begin{equation}
\rho_G^T = \frac{1}{T+1} \sum_{t=0}^T  (\textbf{\emph{U}})^t\Ket{\psi_0}\Bra{\psi_0}(\textbf{\emph{U}}^\top)^t, \label{mixstate2}
\end{equation}
where $\Ket{\psi_0}$ is defined in Def.3 through the weighted Perron-Frobenius operator. $\rho_G^T$ describes a quantum system that consists of a family of equally probable pure states, which are defined by the quantum walk evolution from time step $t = 0$ to $t = T$. Furthermore, we can compute the time-averaged probability $p\{e_d(u,v)\}$ of the quantum walk visiting $e_d(v,u)$
\begin{equation}
p\{e_d(u,v)\}=\rho^T_G((u,v),(u,v)),\label{EqDensityM}
\end{equation}
where $v,u\in V$ and $(u,v)$ indexes $e_d(u,v)$.

\section{Graph Simplification through Commute Time}\label{s3}

The aim of this paper is to develop a novel similarity measure between complete weighted graphs. Directly simulating the evolution of a discrete-time quantum walk on these graphs tends to be elusive, due to the high computational complexity. To overcome this problem, we propose to sparsify the original graphs through the commute time matrix~\cite{DBLP:journals/pami/QiuH07a}.

We first review the concept of commute time. Let $\mathbb{G}$ be a set of complete weighted graphs. Assume $G(V,E)$ is a sample graph from $\mathbb{G}$ with an edge weight function $\mathrm{w}:V\times V \rightarrow \mathbb{R}^+$. If $\mathrm{w} \{u,v\}>0$ or $\mathrm{w} \{v,u\}>0$, there exists an undirected edge $\{u,v\}\in E$ between vertices $v\in V$ and $u\in V$, i.e., $v$ and $u$ are adjacent. Let $A$ denote the adjacency matrix of $G(V,E)$, with entries $A(u,v)=A(v,u)=\mathrm{w}\{u,v\}$. The degree matrix $D$ of $G$ is a diagonal matrix, where each diagonal entry $D(u,u)$ is computed by summing the corresponding row or column of $A$, i.e., $D(u,u) = \sum_v A(u,v)= \sum_v A(v,u)$. Then, the graph Laplacian matrix $L$ is computed by subtracting $A$ from $D$, i.e., $L = D-A$. The spectral decomposition of $L$ is defined as $L=\Phi \Lambda \Phi^T$, where $\Lambda = \mbox{diag}(\lambda_1, \lambda_2, ..., \lambda_n)$ is a $|V| \times |V|$ diagonal matrix with ascending eigenvalues as elements, i.e., $0 = \lambda_1 \leq \lambda_2 \leq ... \leq \lambda_{|V|}$, and $\Phi$ is a $|V| \times |V|$ matrix $\Phi=(\phi_1 | \phi_2 | ... | \phi_{|V|})$ with the corresponding ordered eigenvectors as columns. The hitting time $H(u,v)$ between a pair of vertices $v\in V$ and $u\in V$ of $G(V,E)$ is defined as the expected number of steps of a random walk starting from $u$ to $v$. Similarly, the commute time $C(u,v)$ is computed as the expected number of steps of the walk starting from $u$ to $v$, and then returning to $u$, i.e., $C(u,v)=H(u,v)+H(v,u)$. The commute time $C(u,v)$ can be computed in terms of the unnormalized Laplacian eigendecomposition~\cite{DBLP:journals/pami/QiuH07a}
\begin{equation}
C(u,v) = \sum_{u\in V} D(u,u) \sum_{j=2}^{|V|} \frac{1}{\lambda_j}(\phi_j(u)-\phi_j(v))^2.
\end{equation}
Similarly to~\cite{DBLP:journals/pr/QiuH07b}, we propose to simplify the graph structure by computing a modified commute time spanning tree and reduce the number of edges to $n-1$. More specifically, for a complete weighted graph $G(V,E)\in \mathbb{G}$ and its weighted adjacency matrix $A$, we first compute its commute time matrix $C$ and its associated modified commute time matrix $Q$, where $Q=C\bigodot A$. We then use $Q$ as the new adjacency matrix of $G(V,E)$. Based on Kruskal's method~\cite{Kruskal}, we compute the minimum spanning tree $\mathcal{S}(\mathcal{V},\mathcal{E})$ over $Q$, where $\mathcal{V}=V$.

\noindent\textbf{Discussion} The commute time provides a number of theoretical advantages. \textbf{First}, the commute time amplifies the affinity between pairwise vertices~\cite{Fischer05amplifyingthe} and is robust under the perturbation of the graph structure~\cite{DBLP:journals/pami/QiuH07a}. Thus, the minimum spanning tree $\mathcal{S}$ constructed on the modified commute time matrix can reflect the dominant structural information of the original graph $G$, while yielding a sparser structure~\cite{DBLP:journals/pr/QiuH07b}. \textbf{Second}, as we have stated, the minimum spanning tree $\mathcal{S}$ can reduce the edge number of $G$ to $n-1$. There will be $2(n-1)$ directed edges in $\mathcal{S}$, thus evolving the discrete-time quantum walk on the spanning tree has time complexity $O(n^3)$. This is because the computation of the commute time matrix is based on the eigendecomposition of the normalized Laplacian. Note that this represents a considerable improvement from the original time complexity $O(n^6)$ of simulating the quantum walk on the original graph $G$. \textbf{Third}, the commute time can accommodate the weighted information residing on the graph edges, since it can be computed in terms of the unnormalized Laplacian matrix through the weighted adjacency matrix~\cite{DBLP:journals/pami/QiuH07a}. \textbf{Fourth}, the spanning tree is a simplified structure of the original graph and some weighted edges of the original graph are not encapsulated in the spanning trees. However, since the commute time is defined as the expected step number of the random walk departing from one vertex and returning to the same vertex~\cite{MarMix2008}, i.e., the commute time reflects the integrated effectiveness of all possible paths between pairwise vertices of the original graph~\cite{DBLP:journals/pami/QiuH07a}. Thus, the weighted edges of the spanning tree from the modified commute time matrix can also reflect the weighted information of deleted edges in the original graph. In summary, the minimum spanning tree provides an elegant way to probe the structure of complete weighted graphs using discrete-time quantum walks.

\begin{figure*}[t!]
\vspace{-0pt}
\centering
\subfigure{\includegraphics[width=1\linewidth]{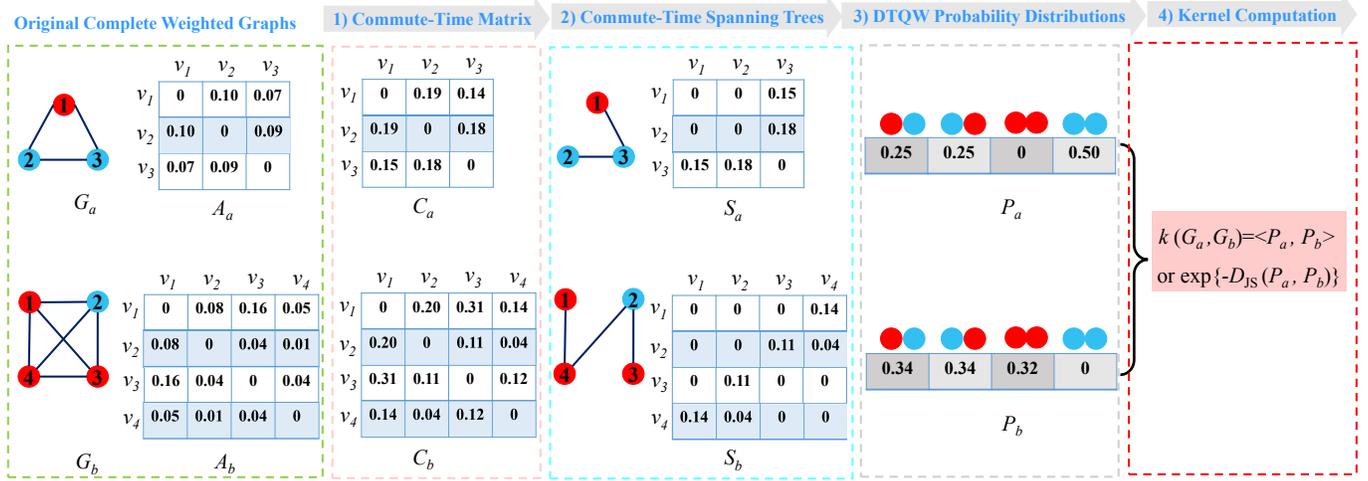}}
\vspace{-15pt}
\caption{An instance of computing the proposed kernels. Let $G_a$ and $G_b$ be two input complete weighted graphs, and $A_a$ and $A_b$ their original weighted vertex adjacency matrices. Note that each color residing on a vertex corresponds to a vertex label. The vertices have the same vertex label if their colors are the same. Specifically, the procedure of computing the proposed kernel between $G_a$ and $G_b$ consists of four steps. (1) The first step computes the commute time matrices, and employs their modified versions $C_a$ and $C_b$ as the new weighted adjacency matrices of $G_a$ and $G_b$, respectively. (2) The second step computes the minimum spanning trees $S_a$ and $S_b$ of $G_a$ and $G_b$ over the modified commute time matrices $C_a$ and $C_b$. Here, the structures of $S_a$ and $S_b$ are sparser than that of their original graphs $G_a$ and $G_b$. (3) The third step probes the spanning trees $S_a$ and $S_b$ in terms of the discrete-time quantum walk, and computes the probability of the quantum walk visiting the directed edges residing on the original edges of the spanning trees. More specifically, this process computes the probability distribution vectors $P_a$ and $P_b$ for $G_a$ and $G_b$, where each vector is spanned by different directed edge labels and each element of the vector corresponds to the sum of the probabilities of the quantum walks visiting the directed edges having the same directed edge label. Here, the combination of each pair of colors corresponds to a directed edge label, where the first and second colors of the combination correspond to the vertex labels of the start and end vertices of a directed edge. (4) The final step computes the kernel based similarity using either a) the dot product or b) the negative exponential of the Jensen-Shannon divergence between the probability distributions $P_a$ and $P_b$.}
\vspace{-10pt}
\label{pipeline_procedure}
\end{figure*}

\section{A Kernel for Complete Weighted Graphs}\label{s4}

In this section we introduce the proposed graph kernel. Given a set of complete weighted graphs, we show that we can associate with each graph a probability distribution over the directed edge labels, which is induced by the time-average probability distribution of the discrete-time quantum walk defined in Section~\ref{s2}. Then, we define two kernels between a pair of graphs in terms of the similarity between the corresponding distributions over the directed edge labels. An instance of computing the proposed kernels between pairwise graphs are exhibited in Fig.\ref{pipeline_procedure}.

\subsection{The Jensen-Shannon Divergence}
The Jensen-Shannon divergence is a non-extensive mutual information measure defined between probability distributions~\cite{DBLP:journals/jmlr/MartinsSXAF09}. Let $\mathcal{P}=(p_1,\ldots,p_m,\ldots,p_M)$ and
$\mathcal{Q}=(q_1,\ldots,q_m,\ldots,q_M)$ be a pair of probability distributions, then the divergence measure
between the distributions is
\begin{align}
D_{JS}(\mathcal{P},\mathcal{Q})& = H_S(\frac{\mathcal{P}+\mathcal{Q}}{2}) - \frac{1}{2} H_S(\mathcal{P}) -
\frac{1}{2} H_S(\mathcal{Q})\nonumber \\
&=-\sum_{m=1}^M \frac{p_m+q_m}{2} \log
\frac{p_m+q_m}{2}\nonumber\\
&+ \sum_{m=1}^M {p_m} \log {p_m}+ \sum_{m=1}^M {q_m} \log
{q_m}.\label{JSDFunctionN}
\end{align}
where $H_S(\mathcal{P})=\sum_{m=1}^M p_m \log p_m$ is the Shannon entropy associated with $\mathcal{P}$. $D_{JS}$ is always negative definite, symmetric, well defined, and bounded, i.e., $0\leq D_{JS}\leq 1$.

\subsection{The Proposed Graph Kernel Over Directed Edge Labels}
Let $\mathbb{G}=\{G_1,\ldots,G_a,\ldots,G_b,\ldots,G_N\}$ denote a set of complete weighted graphs. For a sample graph $G(V,E)\in \mathbb{G}$, we first transform $G$ into a minimum spanning tree $\mathcal{S}(V,\mathcal{E})$ over its modified commute time matrix, based on the method introduced in Section~\ref{s3}. Note that the edges of the commute time spanning tree are attributed with the modified commute time between the corresponding pair of vertices. Let $\mathcal{E}_d$ be the directed edges set of $\mathcal{S}$. Based on Section~\ref{s2}, we simulate the discrete-time quantum walk on the spanning tree $\mathcal{S}$, and we associate with each directed edge $e_d(u,v)\in \mathcal{E}_d$ ($v,u\in V$) the time-average probability $p\{e_d(u,v)\}$ from the quantum walk. Let $L(v)$ be the discrete label associated with the vertex $v\in V$. The directed edge label ${\mathcal{L}}(e_d(u,v))$ of $e_d(u,v)$ is the label union of the start vertex $u$ and end vertex $v$, i.e.,
\begin{equation}
{\mathcal{L}}(e_d(u,v))=\{{L}(u),{L}(v)\}.\label{edgelabele}
\end{equation}
Note that, the resided edge $\{u,v\}\in E$ of $e_d(u,v)\in \mathcal{E}_d$ on the original graph $G(V,E)$ may also have discrete label $L(u,v)$. For this instance, ${\mathcal{L}}(e_d(u,v))$ can be re-written as
\begin{equation}
{\mathcal{L}}(e_d(u,v))=\{{L}(u),{L(u,v)},{L}(v)\}.\label{edgelabelev}
\end{equation}
Thus, we can simultaneously accommodate both discrete vertex and edge labels. Let $\textbf{L}$ be the set of all possible directed edge labels and $l\in \textbf{L}$ is a label. We assign $l$ the probability
\begin{equation}
p(l)=\sum_{v,u\in V}  p\{e_d(u,v)\},\label{DTQWpd}
\end{equation}
where each $e_d(u,v)$ satisfies ${\mathcal{L}}(e_d(u,v))=l$. Note that, for a graph, if the edge label $l$ does not exist, $p(l)=0$.

Let $G_a(V_a,E_a)\in \mathbb{G}$ and $G_b(V_b,E_b) \in \mathbb{G}$ denote a pair of graphs, and $\mathcal{S}_{a}(V_{a},\mathcal{E}_{a})$ and $\mathcal{S}_{b}(V_{b},\mathcal{E}_{b})$ the associated minimum spanning trees. Based on Eq.(\ref{DTQWpd}), we compute the probability distributions over directed edge labels as
$$\mathcal{P}_a=\{p_a(1),\ldots,p_a(l),\ldots,p_a(|\textbf{L}|)\},$$
and
$$\mathcal{P}_b=\{p_b(1),\ldots,p_b(l),\ldots,p_b(|\textbf{L}|)\},$$
associated with the quantum walks on $\mathcal{S}_{a}$ and $\mathcal{S}_{b}$.

Given this setting, we propose to compute the kernel between $G_a$ and $G_b$ in terms of the similarity between $\mathcal{P}_a$ and $\mathcal{P}_b$. We consider two alternative ways of computing this similarity, both resulting in a positive definite kernel.
\\
\noindent
\textbf{1) Measuring Similarity Using the Dot Product:}
Recall that a graph kernel is a positive definite similarity measure which corresponds to the dot product between graphs in an implicit feature space. If we consider $\mathcal{P}_a$ and $\mathcal{P}_b$ as the feature space embeddings of $G_a(V_a,E_a)$ and $G_b(V_b,E_b)$, then the kernel between them can be defined as
\begin{equation}\label{dpkernel}
k_{DP}(G_a,G_b) = \langle \mathcal{P}_a, \mathcal{P}_b \rangle,
\end{equation}
where $\langle.\rangle$ denotes the dot product. The positive definiteness of the kernel trivially follows from Eq.(\ref{dpkernel}). Note that in theory one can employ any standard vector-based kernel measure between the probability distributions $\mathcal{P}_a$ and $\mathcal{P}_b$ as the similarity measure, e.g., the Radial Basis Function (RBF) kernel~\cite{DBLP:journals/neco/ChungKSWL03}, Laplacian kernel~\cite{DBLP:journals/tip/HajiaboliAW12}, etc. However, these vectorial kernels are instances of parametric kernels, i.e., they introduce additional parameters that need to be adjusted in order to achieve the best performance. By contrast, the kernel defined in Eq.(\ref{dpkernel}) is parameter-free and thus does not require the manual setup of additional parameters during the kernel computation

\noindent\textbf{2) Measuring Similarity Using the JS Divergence:}
Based on Eq.(\ref{JSDFunctionN}), the graph kernel between $G_a(V_a,E_a)$ and $G_b(V_b,E_b)$ associated with the Jensen-Shannon divergence is defined as
\begin{align}\label{jskernel}
K_{JS}(G_a,G_b)& = \exp \{-D_{JS}(\mathcal{P}_a,\mathcal{P}_b)\} \nonumber \\
&=\exp \{-H_S(\frac{\mathcal{P}_a+\mathcal{P}_b}{2}) + \frac{H_S(\mathcal{P}_a) + H_S(\mathcal{P}_b)}{2} \} \nonumber \\
&=-\sum_{l\in \textbf{L}} \frac{p_a(l)+p_b(l)}{2} \log
\frac{p_a(l)+p_b(l)}{2}\nonumber\\
&+ \sum_{l\in \textbf{L}} {p_a(l)} \log {p_a(l)}+ \sum_{l\in \textbf{L}}
{p_b(l)} \log {p_b(l)},
\end{align}
where $\frac{\mathcal{P}_a+\mathcal{P}_b}{2}$ is a composite probability distribution and $ \frac{p_a(l)+p_b(l)}{2}$ is computed over the same label $l\in \textbf{L}$. In other words, the similarity between the input graphs is defined in terms of the similarity between the probability distributions over their labels. Note that since these probability distributions are computed over the space of directed edges labels (and not the directed edges themselves), we do not require the two graphs to have the same number of edges.

To see why Eq.(\ref{jskernel}) defines a positive definite kernel, recall that the Jensen-Shannon divergence between probability distributions is a symmetric dissimilarity measure~\cite{majtey2005jensen}. Since the kernel is computed as the negative exponential of the divergence measure, it follows that it is positive definite.

The computational complexity of both kernels is the same. Let $n$ be the maximum number of vertices in a pair of graphs. Then computing $\mathrm{K}_{\mathrm{DP}}$ or $\mathrm{K}_{\mathrm{JS}}$ between these graphs has time complexity $O(n^3)$. In fact, the cost of computing the kernels is dominated by simulating the discrete-time quantum walk on the spanning trees extracted from the original graphs, and this has time complexity $O(n^3)$, as explained in Section~\ref{s3}.

Finally, note that the computation of $\mathrm{K}_{\mathrm{DP}}$ discards any information on a label if this does not appear in both input graphs. For instance, given two graphs $G_a$ and $G_b$, if the label $l$ appears in $G_a$ but not in $G_b$, we have $p_a(l)>0$ and $p_b(l)=0$. As a result, by taking the dot product between the probability vectors, the label $l$ will not influence the resulting kernel value. By contrast, in the computation of $\mathrm{K}_{\mathrm{JS}}$ we make use of all labels, including those that appear in only one of the two input graphs. This in turn suggests that $\mathrm{K}_{\mathrm{JS}}$ may potentially reflect richer graph characteristics than $\mathrm{K}_{\mathrm{DP}}$ for some instances.

\subsection{Discussions and Related Works}
\begin{table*}
\centering {
% \tiny
% \scriptsize
\footnotesize
\caption{Summary statistics for the selected graph datasets}\label{Comparison}
\vspace{-10pt}
\begin{tabular}{|c||c||c||c||c|}

  \hline
  % after \\: \hline or \cline{col1-col2} \cline{col3-col4} ...
 ~Property ~                                    & ~\textbf{The Proposed Kernels}  ~      &~QJSK~\cite{DBLP:conf/gbrpr/Bai0RZH15}~    & ~QJSKT~\cite{DBLP:journals/prl/Bai16}~ &  ~QEMK~\cite{DBLP:conf/iciap/BaiZR0H15}~ \\ \hline \hline

 ~Permutation Invariant~                        & ~$\mathrm{Yes}$~  &~ $\mathrm{No}$~  & ~$\mathrm{No}$~ &  ~$\mathrm{Yes}$~ \\  \hline

 ~Time Complexity for N Graphs~        & ~$O(Nn^3+N^2n)$~  &~ $O(N^2n^6)$~    & ~$O(N^2n^3)$~   &  ~$O(Nn^6+N^2n^4)$~ \\ \hline

 ~Time Complexity for Pairwise Graphs~ & ~$O(n^3)$~        &~ $O(n^6)$~       & ~$O(n^3)$~      &  ~$O(n^6)$~ \\ \hline

 ~Positive Definite~                            & ~$\mathrm{Yes}$~  &~ $\mathrm{Yes}$~ & ~$\mathrm{Yes}$~&  ~$\mathrm{No}$~    \\ \hline

 ~Accommodate Attributed Graphs~       & ~$\mathrm{Yes}$~  &~ $\mathrm{No}$~  & ~$\mathrm{No}$~ &  ~$\mathrm{No}$~ \\  \hline

 ~Accommodate Edge Weights~       & ~$\mathrm{Yes}$~  &~ $\mathrm{No}$~  & ~$\mathrm{No}$~ &  ~$\mathrm{No}$~ \\  \hline

\end{tabular}
}
\vspace{-15pt}
\end{table*}

The proposed graph kernels are related to the quantum Jensen-Shannon kernel (QJSK)~\cite{DBLP:conf/gbrpr/Bai0RZH15} and its faster variant (QJSKT)~\cite{DBLP:journals/prl/Bai16}, as well as the quantum edge-based matching kernel (QEMK)~\cite{DBLP:conf/iciap/BaiZR0H15}. All these kernels use discrete-time quantum walks to probe the graph structure. Moreover, the proposed kernels are also similar to the QJSKT kernel, which is also based on the spanning trees extracted from original graph through the commute time. Thus, the QJSKT kernel for a pair of graphs both having $n$ vertices also has time complexity $O(n^3)$. This is significantly more efficient than the QJSK kernel that requires time complexity $O(n^6)$. However, there are five significant theoretical differences between the proposed kernels and these kernels. These differences are listed in Table~\ref{Comparison} and discussed as follows.

First, as we have stated in Section~\ref{s3}, the minimum spanning tree, which is extracted from the original graph through the commute time, can reflect weight information residing on the edges of the original graph. Thus, unlike the QJSK kernel between original graphs, the proposed kernels between commute time spanning trees not only overcome the shortcoming of inefficiency, but also encapsulate the edge weight information of original graphs through the proposed weighted Perron-Frobenius operator.

Second, like the proposed kernels, the QJSKT kernel is also defined on the commute time spanning tree and requires the same time complexity $O(n^3)$ for a pair of graphs. The computation of the initial quantum state of the QJSKT kernel is based on the unweighted Perron-Frobenius operator rather then the weighted operator. This is similar to the QJSK kernel. Thus, unlike the proposed kernels, the QJSKT kernel ignores edge weight information of the spanning trees, i.e., it does not reflect the edge weight information of the original graphs.

Third, unlike the QJSKT and QJSK kernels, the proposed kernels are based on the dot product and the classical Jensen-Shannon divergence between probability distributions over directed edge labels, respectively. These distributions, on the other hand, correspond to the diagonals of the density operators of the discrete-time quantum walks. Since the labels for the proposed kernels are computed by taking the union of the corresponding vertex/edge labels, the proposed kernels can also accommodate vertex/edge attributed graphs. By contrast, the QJSKT and QJSK kernels are based on the quantum Jensen-Shannon divergence~\cite{DBLP:journals/prl/Bai16} between the density operators associated with the discrete-time quantum walks. This in turn requires computing their eigendecomposition, which has time complexity $O(n^3)$. On the other hand, both the dot product and the classical Jensen-Shannon divergence between a pair of probability distributions have time complexity $O(n)$. As a result, for a set of $N$ graphs each having $n$ vertices, the proposed kernels only have time complexity $O(Nn^3+N^2n)$. This is significantly more efficient than the QJSKT kernel, which has time complexity $O(N^2n^3)$.

Fourth, both the QJSKT kernel and the QJSK kernel require computing a composite density operator from the pair of density operator under comparisons. However, when computing this composite density operator both kernels do not take into account the correspondences between the vertices of the directed line graphs, i.e., the directed edges residing on the original graph edges. As a result, neither kernel is permutation invariant. By contrast, in the proposed kernels we overcome the problem by comparing probability distributions over the space of the directed edge labels. The proposed kernels are thus permutation invariant, since permuting the directed edge order (i.e., the adjacency matrix of the corresponding directed line graph) does not change its set of directed edge labels. In other words, our kernels gives a more precise kernel based similarity measure than the QJSKT and QJSK kernels.

Finally, like the QJSK kernel, the QEMK kernel also involves evolving a discrete-time quantum walk, but on the original graph. Thus, the QEMK kernel also suffers from the inefficiency of simulating quantum walks on the original graph. Moreover, computing the QEMK kernel between a pair of graphs requires aligning their vertices through their depth-based representations~\cite{DBLP:journals/prl/Bai16}. Since this alignment step is not guaranteed to be transitive, the resulting kernel is not positive definite (\textbf{\emph{pd}})~\cite{DBLP:conf/icml/FrohlichWSZ05}. By contrast, the proposed kernels are \textbf{\emph{pd}}.

\begin{figure*}[t!]
\vspace{-0pt}
\centering
\subfigure{\includegraphics[width=1\linewidth]{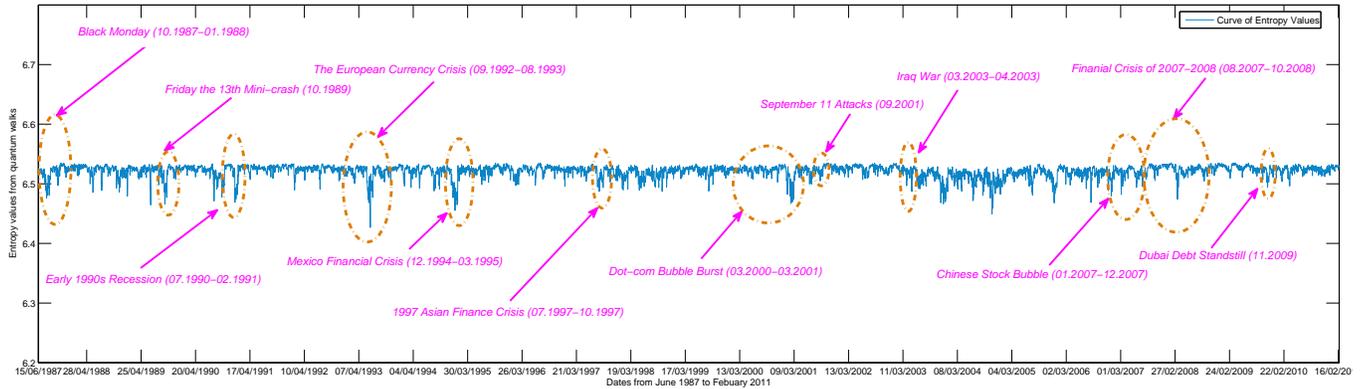}}
\vspace{-20pt}
\caption{Shannon entropy versus time for time-varying financial networks.}\label{f:entropy1}
\vspace{-0pt}
\end{figure*}
\begin{figure*}[t!]
%\vspace{-10pt}
\centering
\subfigure[QK kPCA Embeddings]{\includegraphics[width=0.192\linewidth]{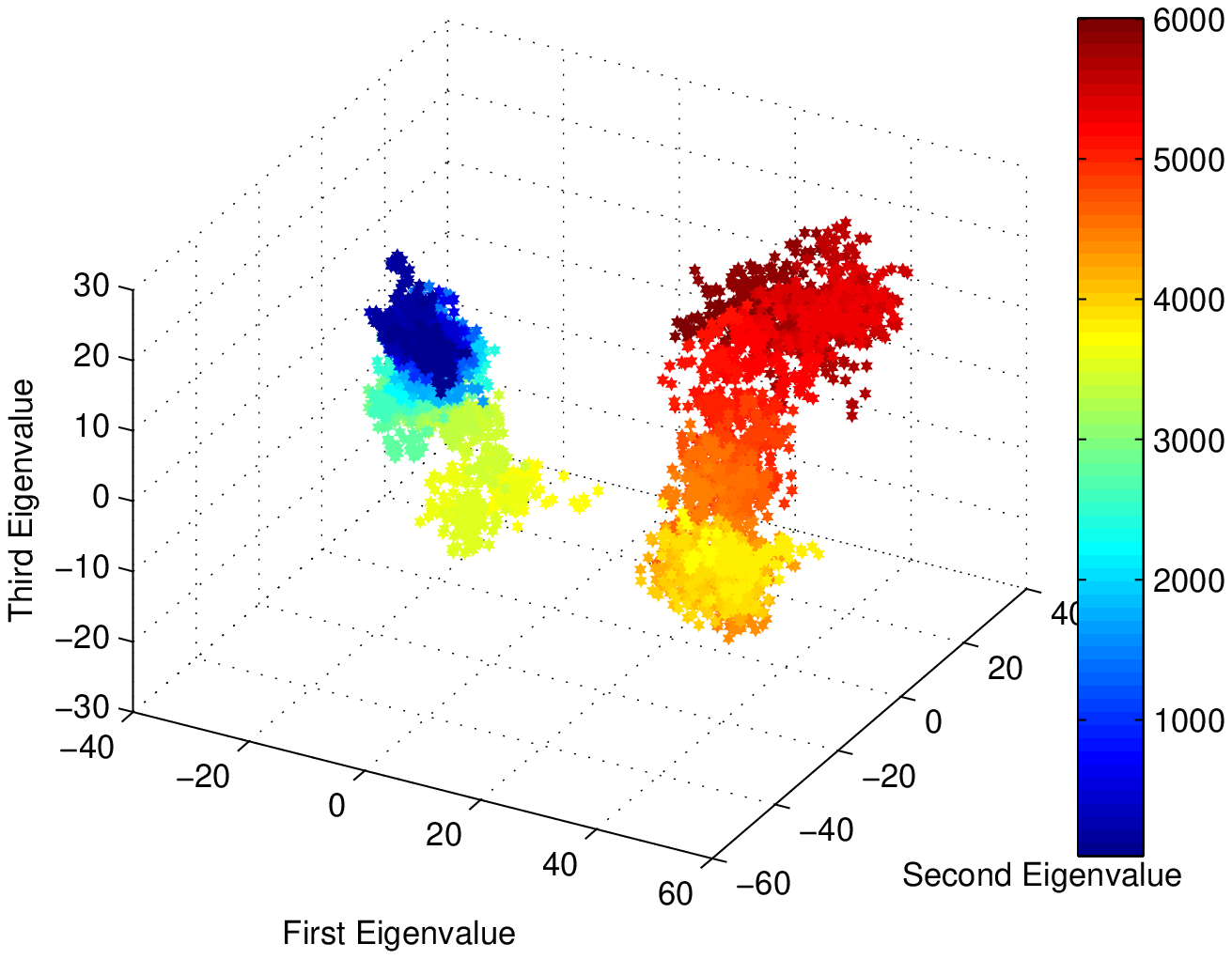}}
\subfigure[WLSK kPCA Embeddings]{\includegraphics[width=0.192\linewidth]{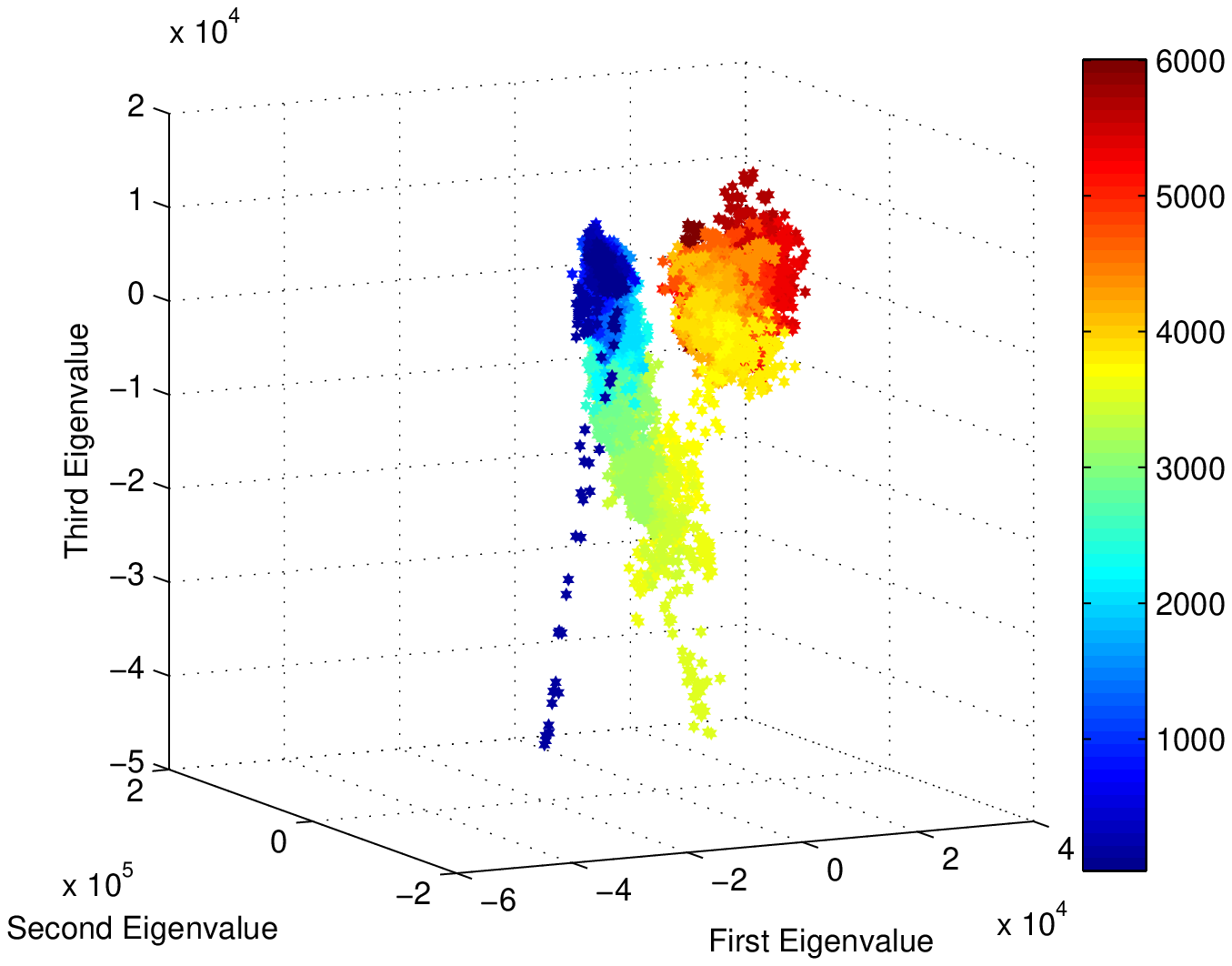}}
\subfigure[JSGK kPCA Embeddings]{\includegraphics[width=0.192\linewidth]{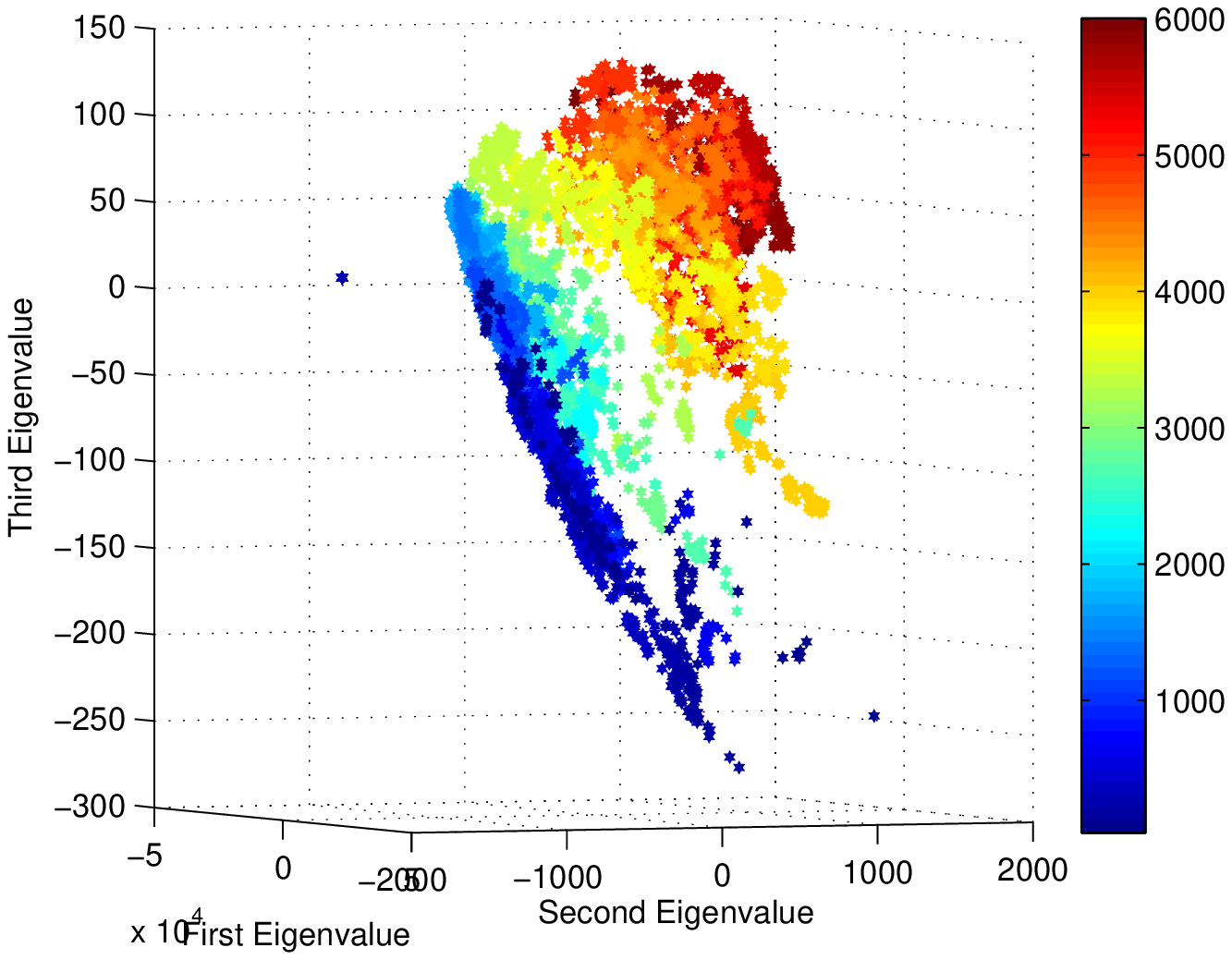}}
\subfigure[QJSK kPCA Embeddings]{\includegraphics[width=0.192\linewidth]{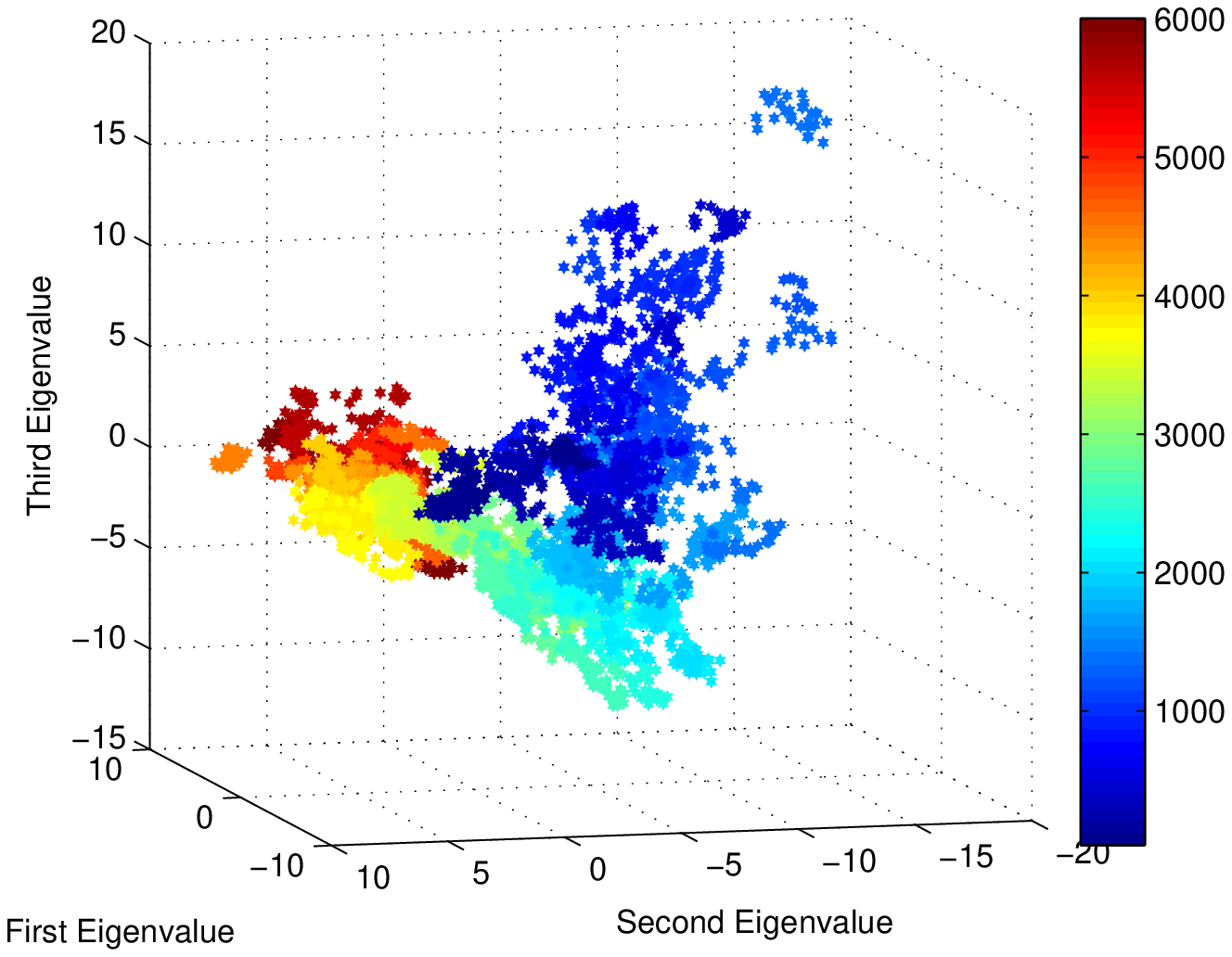}}
\subfigure[FLGK kPCK Embeddings]{\includegraphics[width=0.192\linewidth]{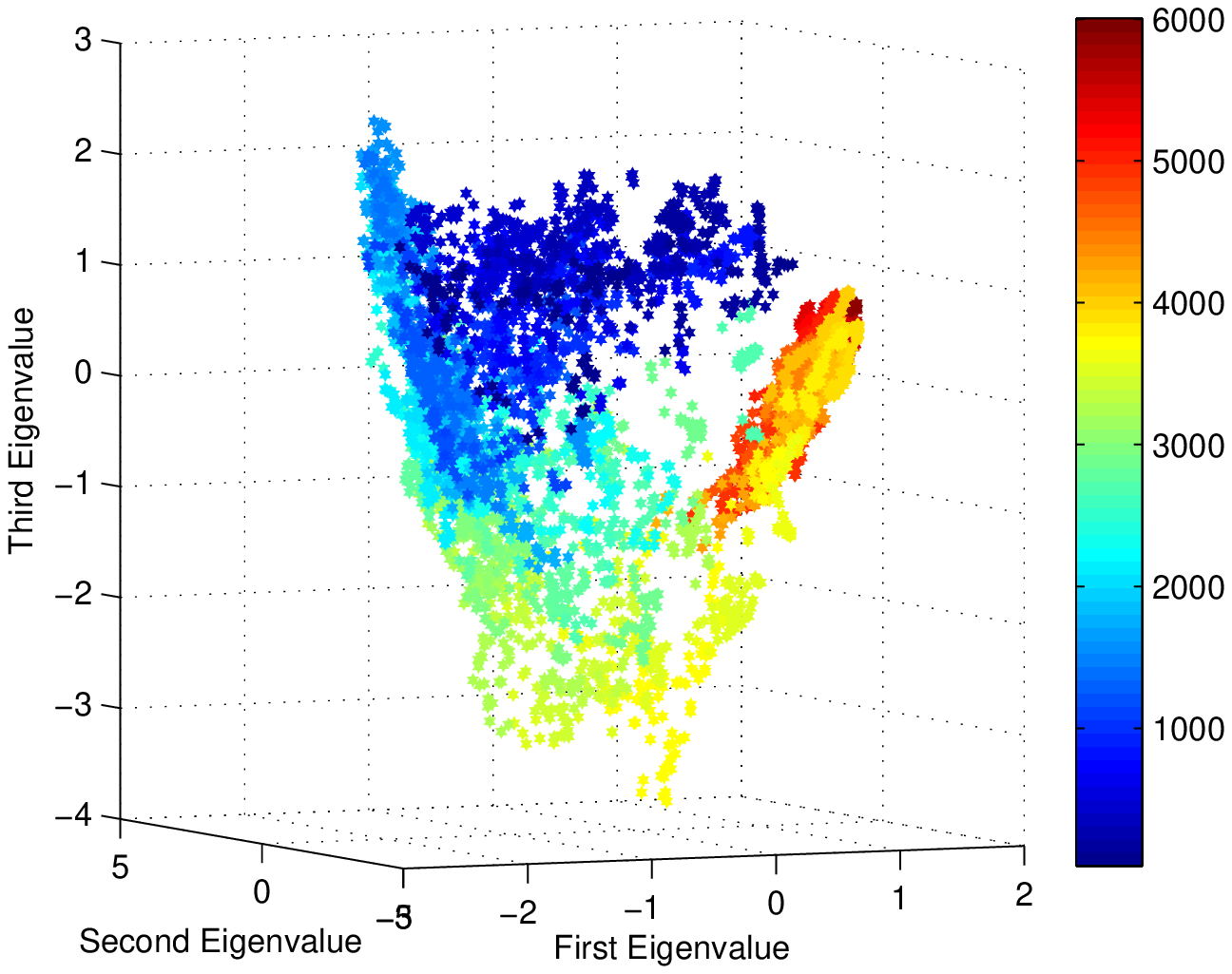}}
\vspace{-10pt}
\caption{(Color online) Path of financial networks over 5976 days based on kPCAs of different graph kernels.}\label{allembeddings}
\vspace{-10pt}
\end{figure*}
\section{Experiments}\label{s5}

As an example of a complete weighted network, we consider the case of time-varying financial networks. The \textbf{NYSE dataset} consists of a series of networks for each trading day abstracted from the closing price of stocks from the New York Stock Exchange (NYSE) database~\cite{silva2015modular}, which consists of 3799 stocks and their associated daily prices. The stock prices were obtained from the Yahoo financial dataset (http://finance.yahoo.com)~\cite{ye2015thermodynamic}. To extract the network representations, we select 347 stocks that were traded from January 1986 to February 2011, i.e., for a total of 6004 days. We select a period of 28 days as the time window to analyse the similarity of the closing prices of different stocks and we move this window along the 6004 trading days to construct a time varying sequence of stock prices. More specifically, each chronological window of the sequence encapsulates a time series of stock return values over a corresponding period of 28 days. For each time window, i.e., each trading day, we represent the trades between each pair of stock as a network where the connection weight between two stocks is the Euclidean distance between their time series. The resulting structure is a time-varying network on 347 vertices over 5976 days. Note that each network is a complete weighted graph, and each vertex label corresponds to a corresponding stock name (the edges do not have discrete labels). To our knowledge, the aforementioned existing state-of-the-art R-convolution graph kernels cannot directly accommodate this kind of structures, since none of them can deal with complete weighted graphs.

\subsection{Evaluations of the Network Entropy}

We commence by investigating the Shannon entropies associated with the directed edge label probability distributions
induced by the discrete-time quantum walks. These entropies play a significant role in determining the kernel performance. Specifically, we explore the evolutionary behavior of the NYSE stock market by computing the entropy of the time varying financial networks at each time step. This allows us to investigate whether the evolutionary behaviour
can be understood through  the variation of the network  entropy, i.e., we aim to analyze whether abrupt changes in  network structure or different evolutionary epochs can be characterized by the entropy measure. Note that, we evolve the quantum walks from $t=1$ to $t=25$, i.e., we set the maximum $t$ as $T=25$, because the Shannon entropy computed through the quantum walk tends to reach a limiting value when $T\geq 25$. The results are shown in Fig. \ref{f:entropy1}, where the x-axis represents the date (time) and the y-axis represents entropy values. Fig. \ref{f:entropy1} indicates that most of the significant fluctuations in the entropy time series correspond to different financial crises, e.g., Black Monday~\cite{blackmonday}, the Friday the 13th minicrash~\cite{BOOK2}, the Dot-com Bubble Burst, the Mexico Financial Crisis, the Iraq War, and the Subprime Crisis~\cite{subprime}, etc. This is because the time-varying financial network experiences dramatic structural changes when a financial crisis occurs. For instance, some significant Internet companies that led to a rapid increase of both market confidence and stock prices were identified during the period of the Dot-com Bubble~\cite{anderson2010speculative}. This noticeably modified the subsequent relations between stock, and this phenomenon can be captured by examining the variation of the Shannon entropy values.

Note that, although Fig. \ref{f:entropy1} demonstrates that the entropy is effective in detecting the extreme events in the financial network evolution, the time series is one dimensional and hence overlooks information concerning detailed changes in network structure. By contrast, the proposed quantum walk kernels can map the network structures into a high dimensional space by kernelizing the entropy and better preserve structural information contained in the networks.

\subsection{Quantum Kernel Embeddings from kPCA}

In this subsection, we explore the effectiveness of the proposed quantum kernels on the NYSE dataset. To this end, we apply the new kernels to the time-varying financial networks with the objective of analyzing whether abrupt changes in network evolution can be distinguished. Note that in the remainder of this subsection we choose to show only the results for the kernel based on the dot product, as we observed that these were very similar to the ones computed using the Jensen-Shannon divergence and the computation is faster.

We commence by setting $T=25$ for the evolution of the required discrete-time quantum walks. We perform kernel Principle Component Analysis (kPCA)~\cite{witten2011data} on the kernel matrix of the financial networks (from January 1986 to February 2011) using the proposed kernel, and we embed the networks into a 3-dimensional component space. The results from the kPCA are visualized using the first three principal components and are exhibited in Fig. \ref{allembeddings}(a). Moreover, we compare the proposed quantum kernel with four state-of-the-art kernels, i.e., the Weisfeiler-Lehman subtree kernel (WLSK)~\cite{DBLP:journals/jmlr/ShervashidzeVPMB09}, the Jensen-Shannon graph kernel (JSGK)~\cite{DBLP:journals/jmiv/BaiH13}, the quantum Jensen-Shannon kernel (QJS)~\cite{DBLP:journals/pr/Bai0TH15} and the feature space Laplacian graph kernel (FLGK)~\cite{DBLP:conf/nips/KondorP16}. Note that, the WLSK is an instance of the R-convolution kernel, and it cannot accommodate either complete weighted graphs or weighted graphs. Thus, we apply the WLSK kernel to the transformed commute time spanning trees and ignore the weight information on the tree edges. The JSGK kernel is a mutual information kernel associated with the steady state random walk, the QJS kernel is a quantum kernel associated with the continuous-time quantum walk, and the FLGK kernel is a global kernel associated with the Laplacian matrix. Since all the JSGK, QJS and FLGK kernels can accommodate edge weight, we directly apply these kernels to the original financial networks. Moreover, since each vertex label appears just once for each network, we establish the required correspondences between a pair of networks through the vertex labels for the JSGK and QJS kernels. We also perform kPCA on the resulting kernel matrices and embed the networks into a 3-dimensional principal component space. The embedding results for the WLSK, JSGK, QJS and FLGK kernels are visualized in Fig. \ref{allembeddings}(b), \ref{allembeddings}(c), \ref{allembeddings}(d) and \ref{allembeddings}(e). The plots in Fig. \ref{allembeddings} indicate the paths of the time-varying financial networks in the different kernel spaces over 5976 trading days. The color bar beside each plot represents the date in the time series. It is clear that the embedding given by the proposed kernel shows a better manifold structure. Moreover, an interesting phenomenon in Figs.\ref{allembeddings}(a) and \ref{allembeddings}(b) is that networks from Jan 1986 to Feb 2011 are well divided into two clusters. The clusters found by our kernel are better separated.

\begin{figure}
\centering
\subfigure[Black Monday]{\includegraphics[width=0.49\linewidth]{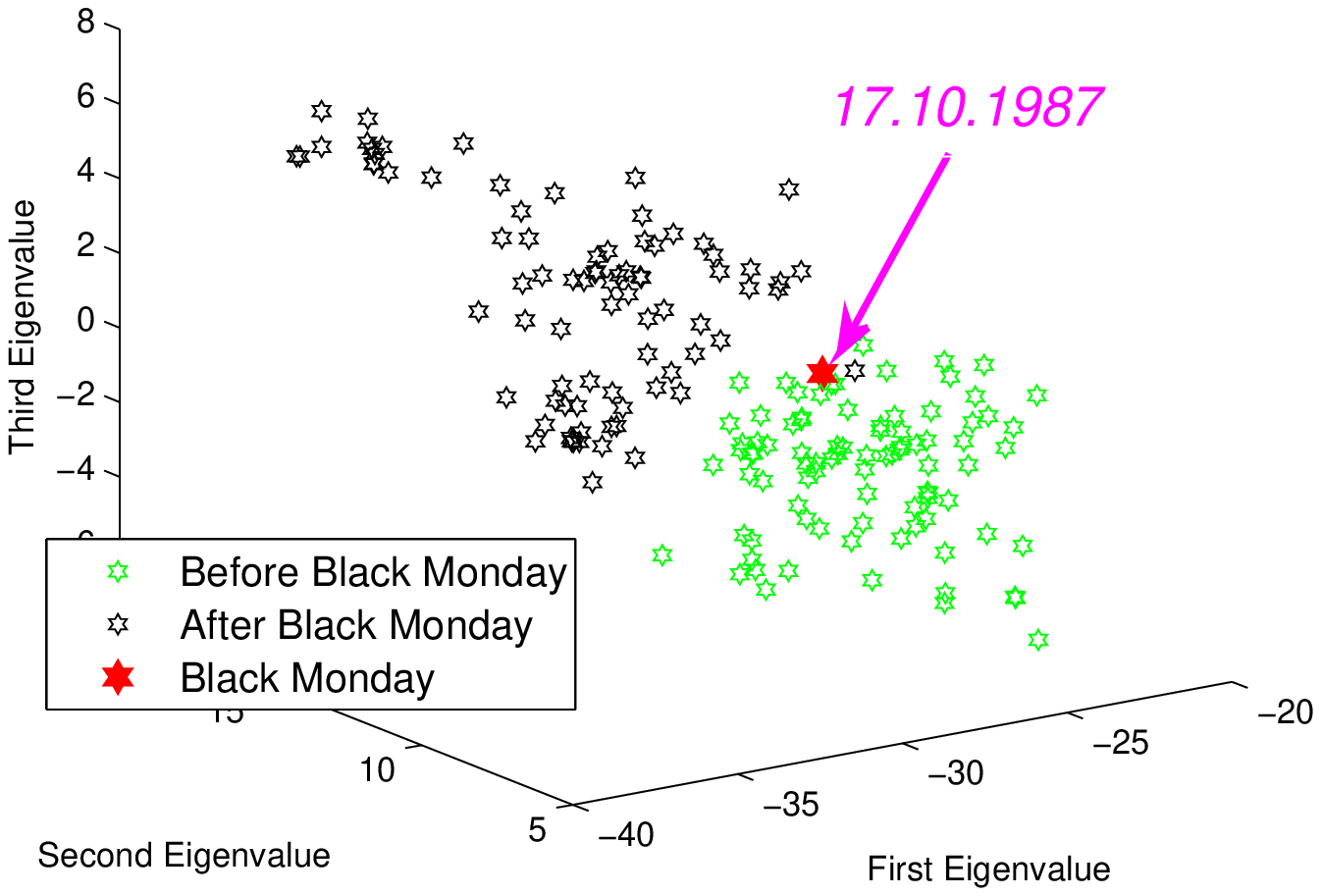}}
\subfigure[Dot-com Bubble]{\includegraphics[width=0.49\linewidth]{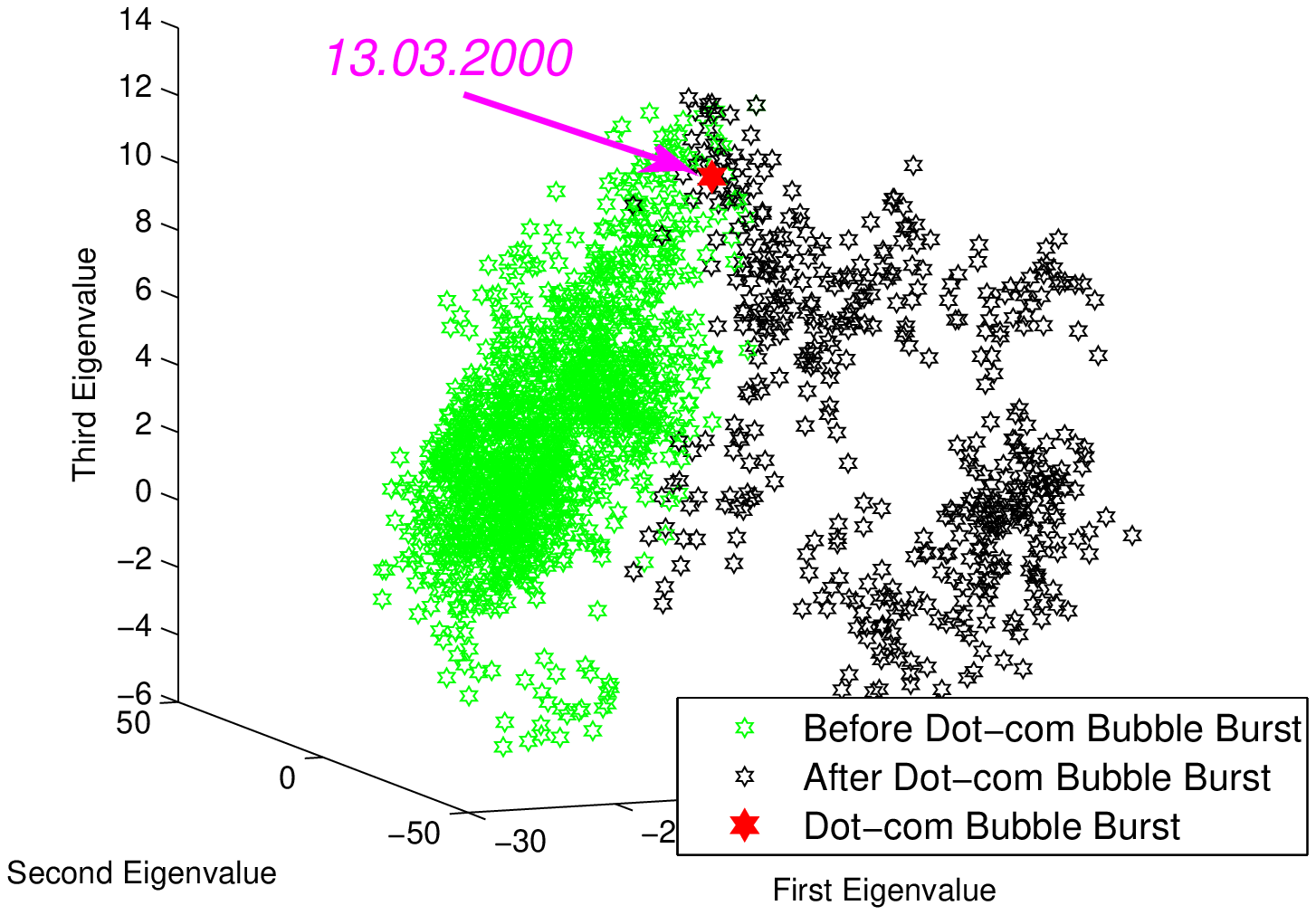}}
\subfigure[Enron Incident]{\includegraphics[width=0.49\linewidth]{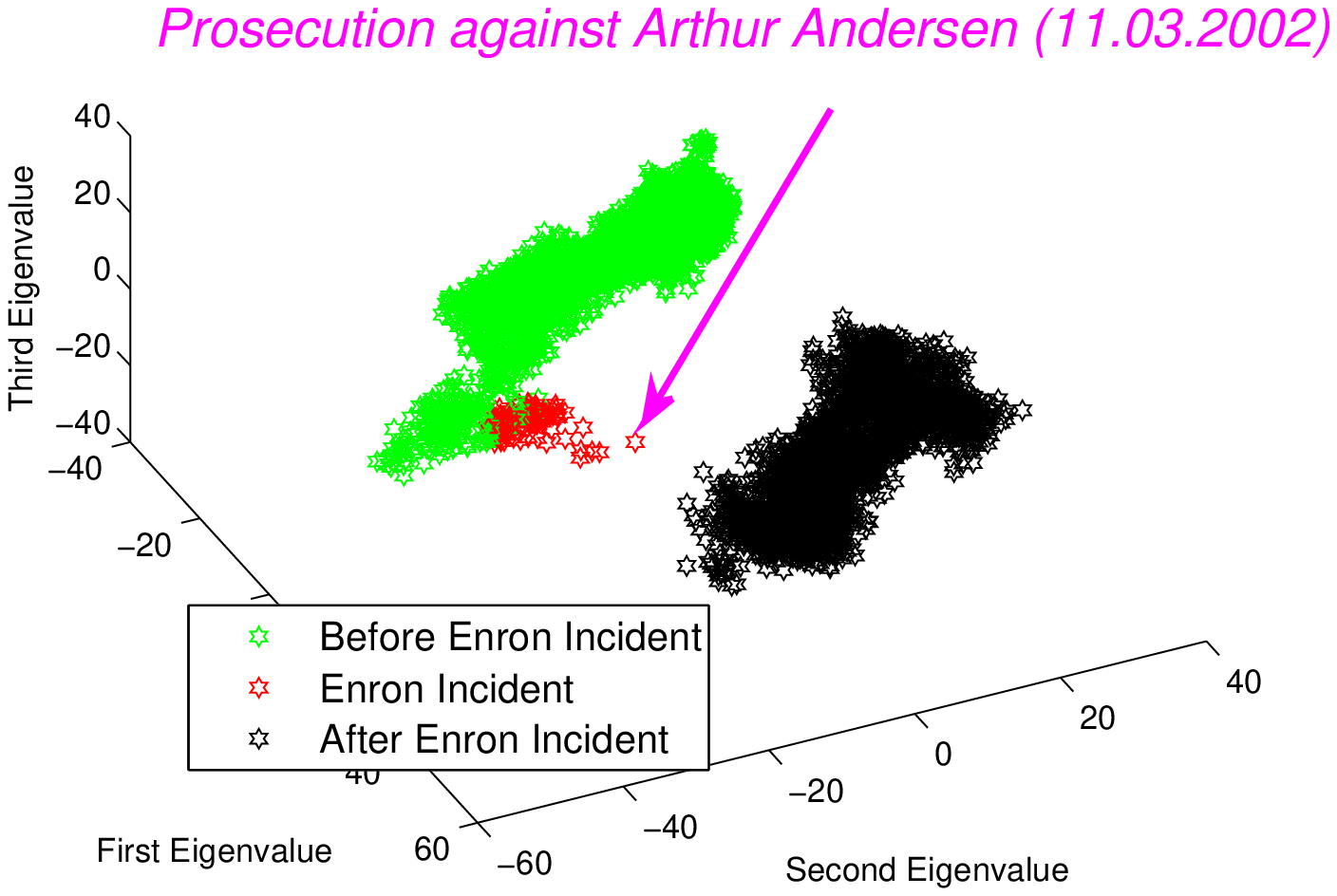}}
\subfigure[Subprime Crisis]{\includegraphics[width=0.49\linewidth]{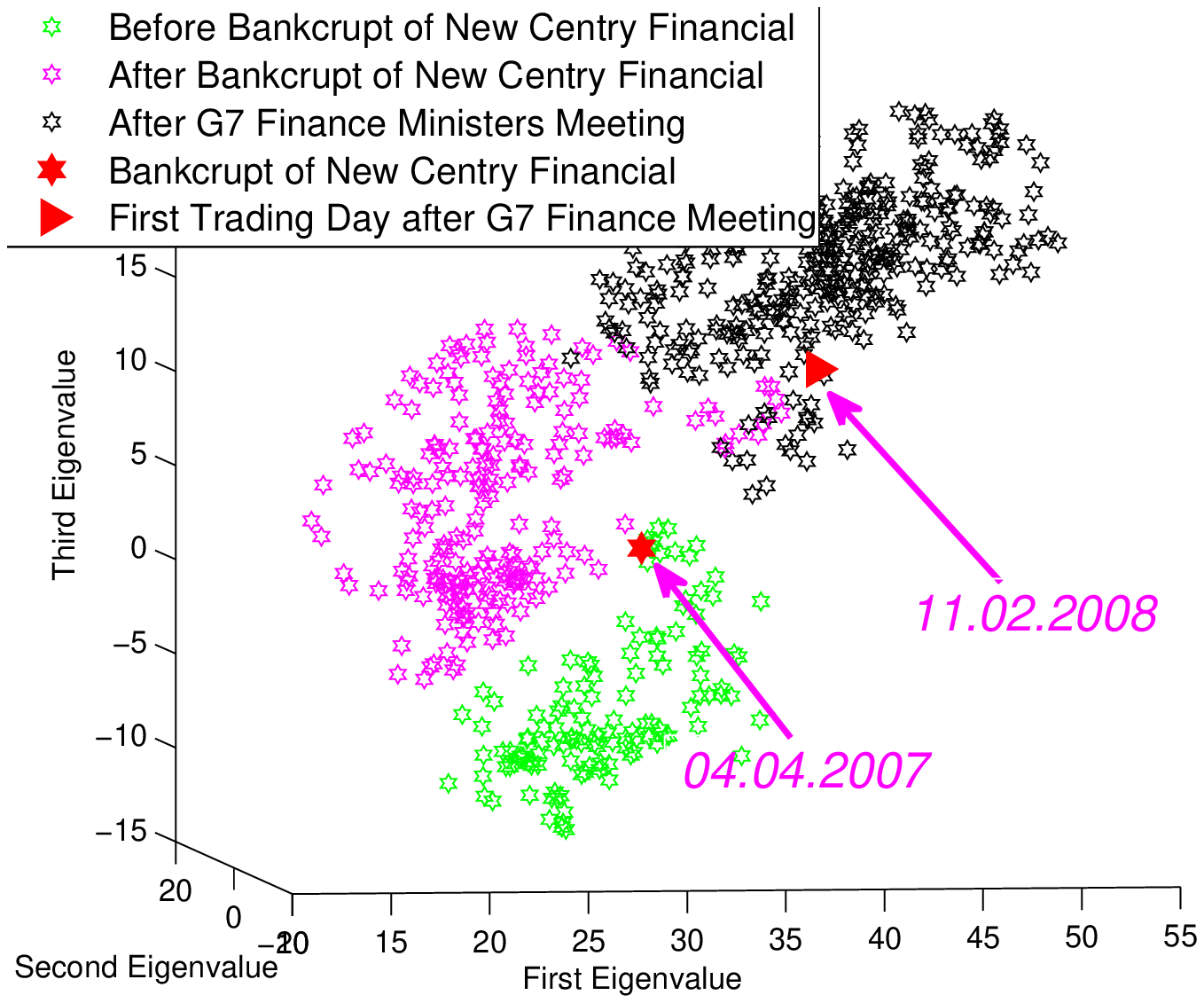}}
\vspace{-10pt}
\caption{The 3D embedding of the financial networks during different finance crises based on kPCA from QK (for Commute Time Spanning Trees).}\label{embeddings}
\vspace{-10pt}
\end{figure}

To take this study one step further, we show in detail the kPCA embeddings during four different financial crisis periods. Specifically, Fig. \ref{embeddings}(a) corresponds to the Black Monday period (\emph{\textbf{from 15th Jun 1987 to 17th Feb 1988}}), Fig. \ref{embeddings}(b) to the Dot-com Bubble period (\emph{\textbf{from 3rd Jan 1995 to 31st Dec 2001}}), Fig. \ref{embeddings}(c) to the Enron Incident period (\emph{\textbf{the red points, from 16th Oct 2001 to 11th Mar 2002}}), and Fig. \ref{embeddings}(d) to the Subprime Crisis period (\emph{\textbf{from 2nd Jan 2006 to 1st Jul 2009}}). Fig. \ref{embeddings}(a) indicates that Black Monday (\emph{\textbf{17th Oct, 1987}}) is a crucial financial event, as the network embedding points before and after the event are divided into two clusters. Similarly, Fig. \ref{embeddings} indicates that the Dot-com Bubble Burst (\emph{\textbf{13rd Mar, 2000}}) in Fig. \ref{embeddings}(b), the Enron Incident period (\emph{\textbf{from 2nd Dec 2001 to 11th Mar 2002}}) in Fig. \ref{embeddings}(c),  the New Century Financial crisis (\emph{\textbf{4th April, 2007}}) and the First Trading Day after G7 Finance Meeting (\emph{\textbf{2nd Nov, 2008}}) in Fig. \ref{embeddings}(d) are also crucial financial events. The network embedding points before and after these events are separated into distinct clusters. Moreover, points corresponding to the crucial events are midway between the two clusters. Another interesting feature in Fig. \ref{embeddings}(c) is that the networks between 1986 and 2011 are separated by the Prosecution against Arthur Andersen (\emph{\textbf{3rd Nov, 2002}}). The prosecution is closely related to the Enron Incident. As a result, the Enron Incident can be seen as a watershed at the beginning of 21st century, that significantly distinguishes the financial networks of the 21st and 20th centuries.

\begin{figure}
\centering
\subfigure[Black Monday]{\includegraphics[width=0.49\linewidth]{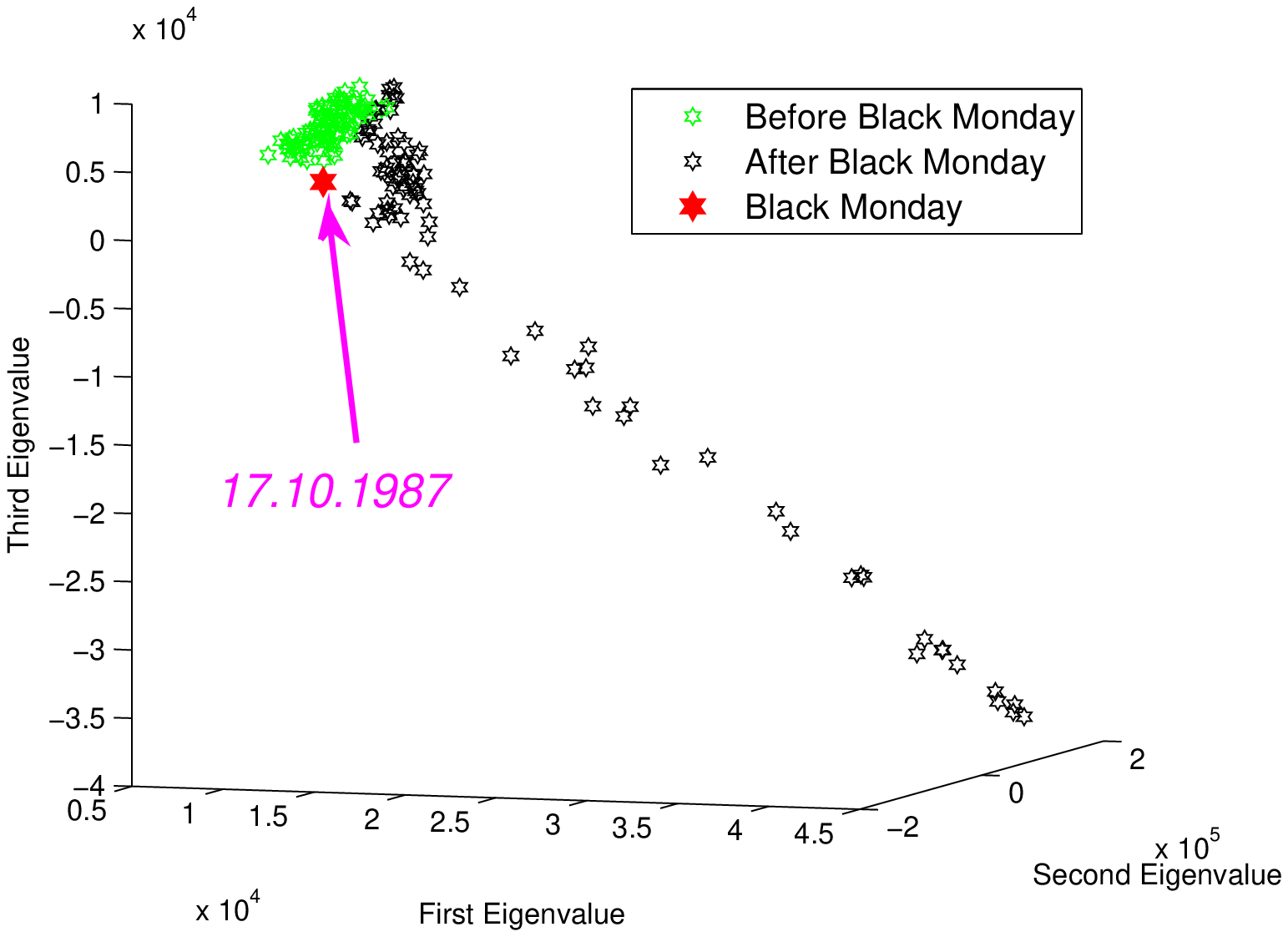}}
\subfigure[Dot-com Bubble]{\includegraphics[width=0.49\linewidth]{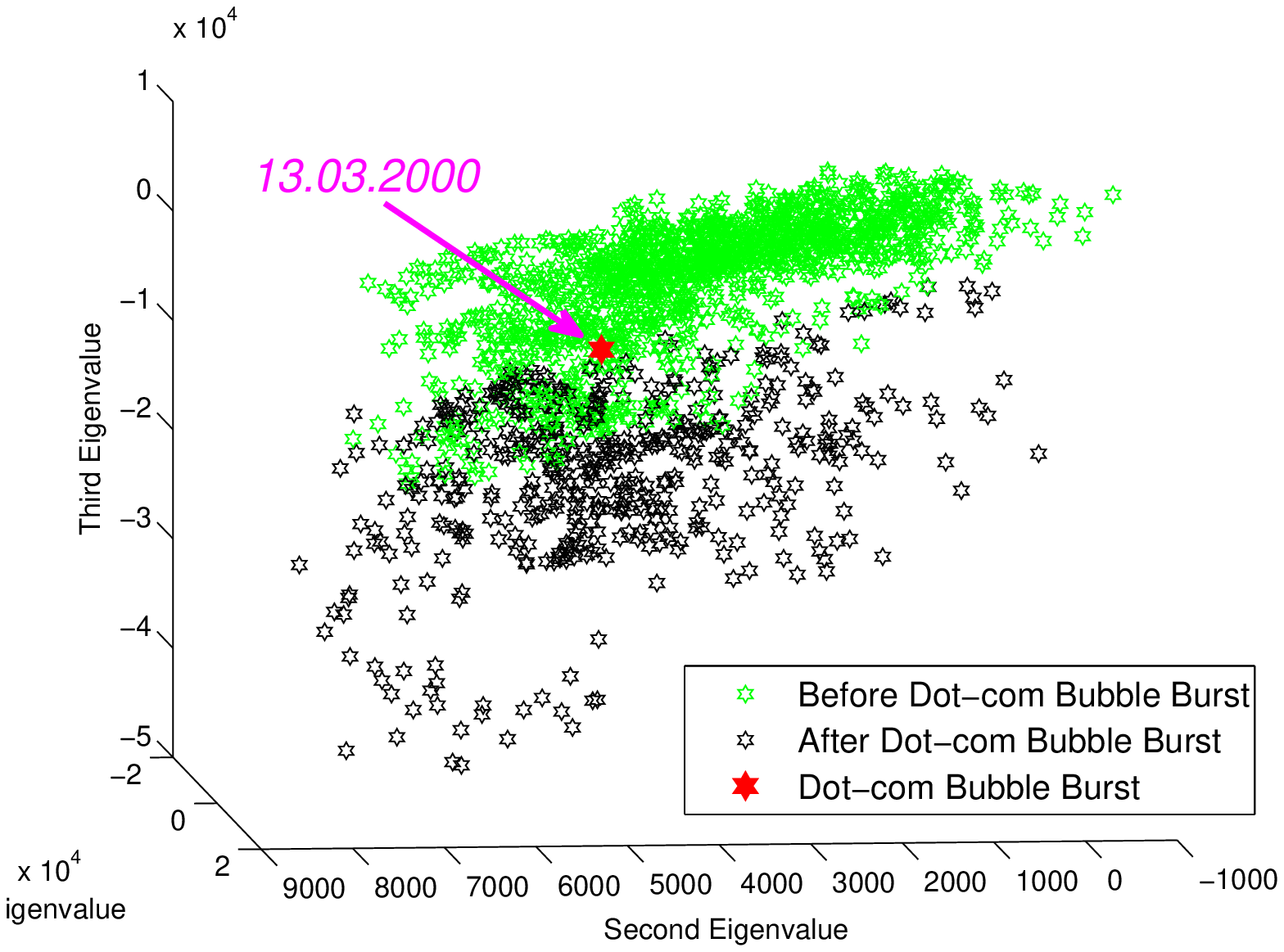}}
\subfigure[Enron Incident]{\includegraphics[width=0.49\linewidth]{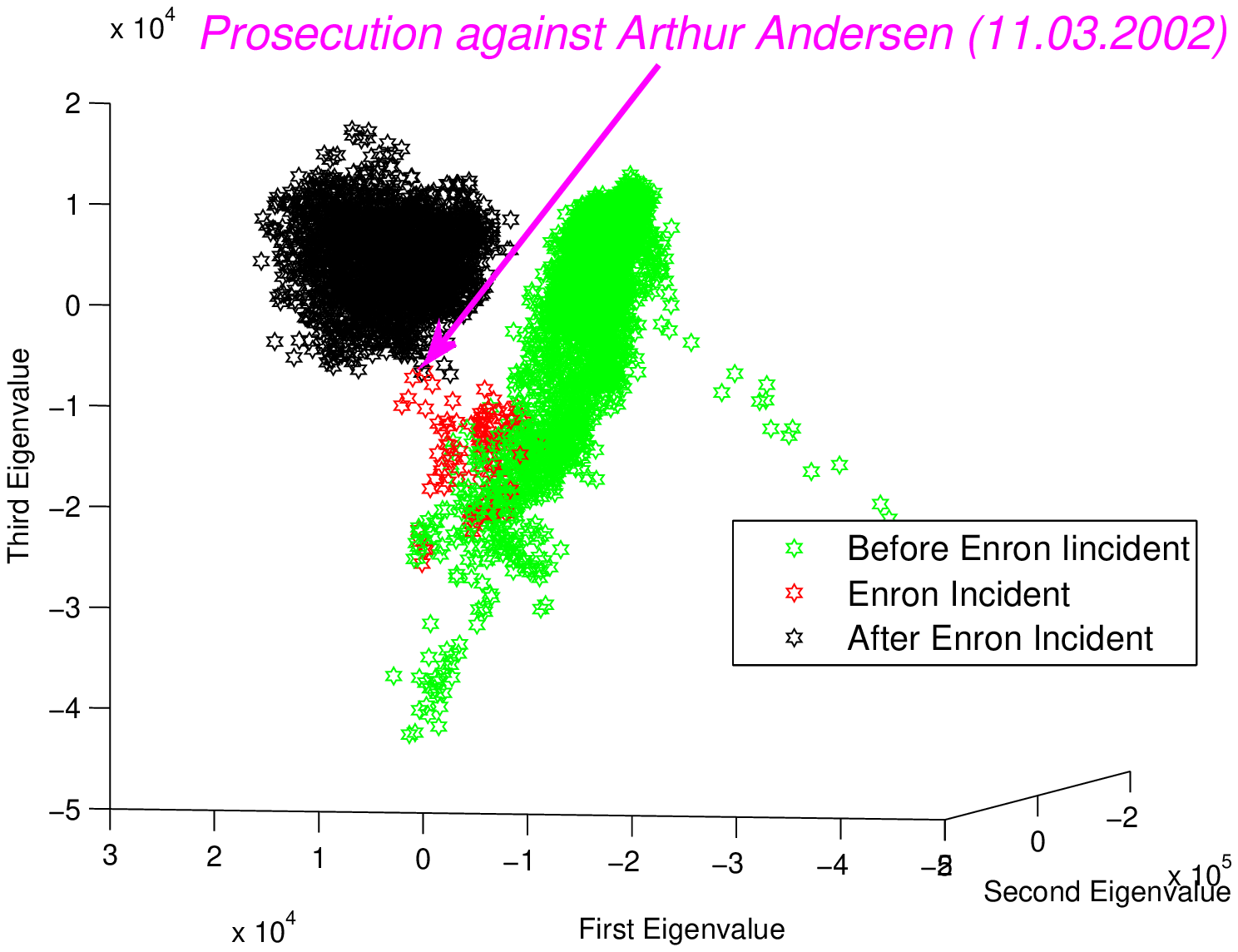}}
\subfigure[Subprime Crisis]{\includegraphics[width=0.49\linewidth]{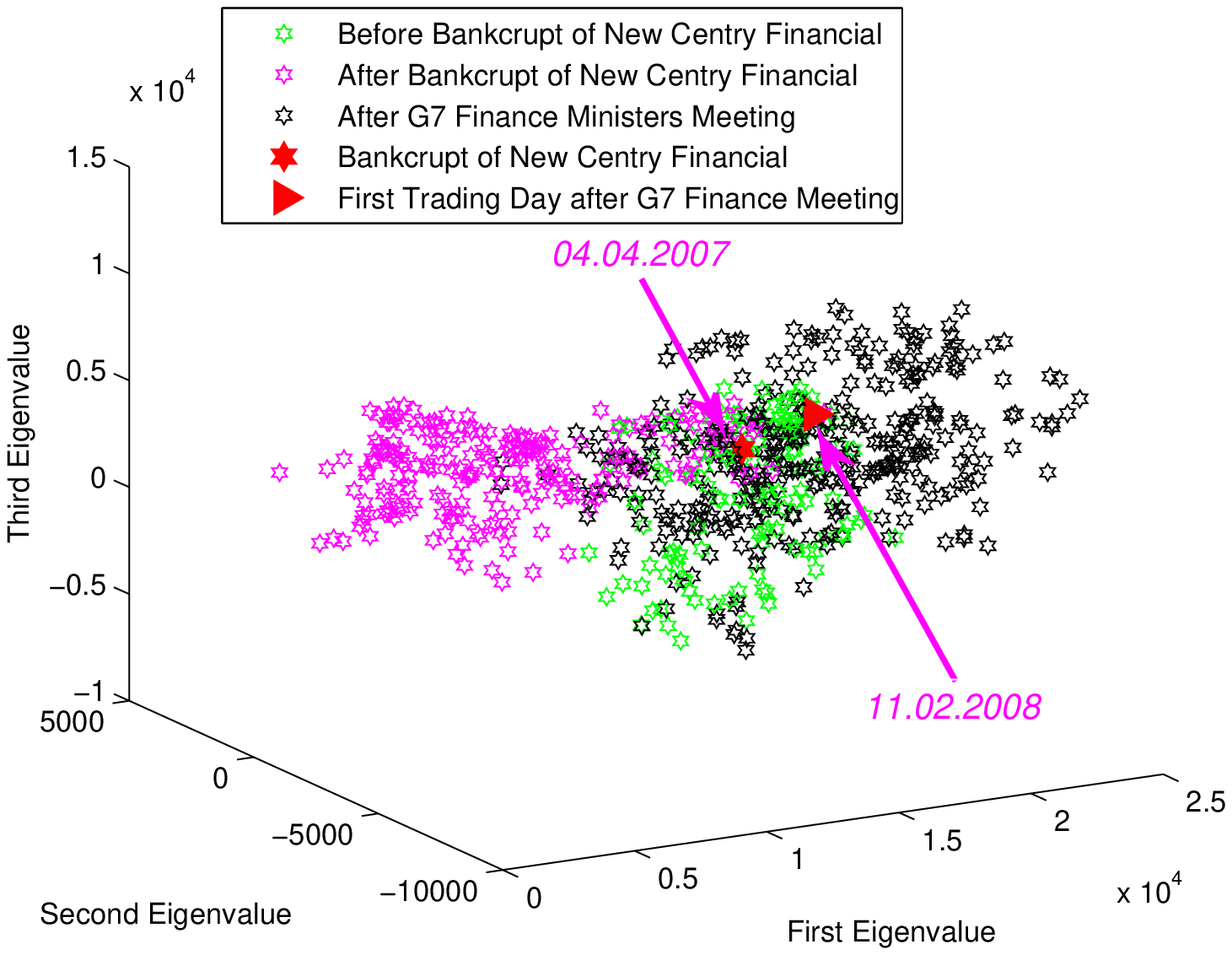}}
\vspace{-10pt}
\caption{The 3D embedding of the financial Networks during Different Finance Crises based on kPCA from WLSK.} \label{embeddingsCompareWL}
\vspace{-10pt}
\end{figure}

\begin{figure}
\centering
\subfigure[Black Monday]{\includegraphics[width=0.49\linewidth]{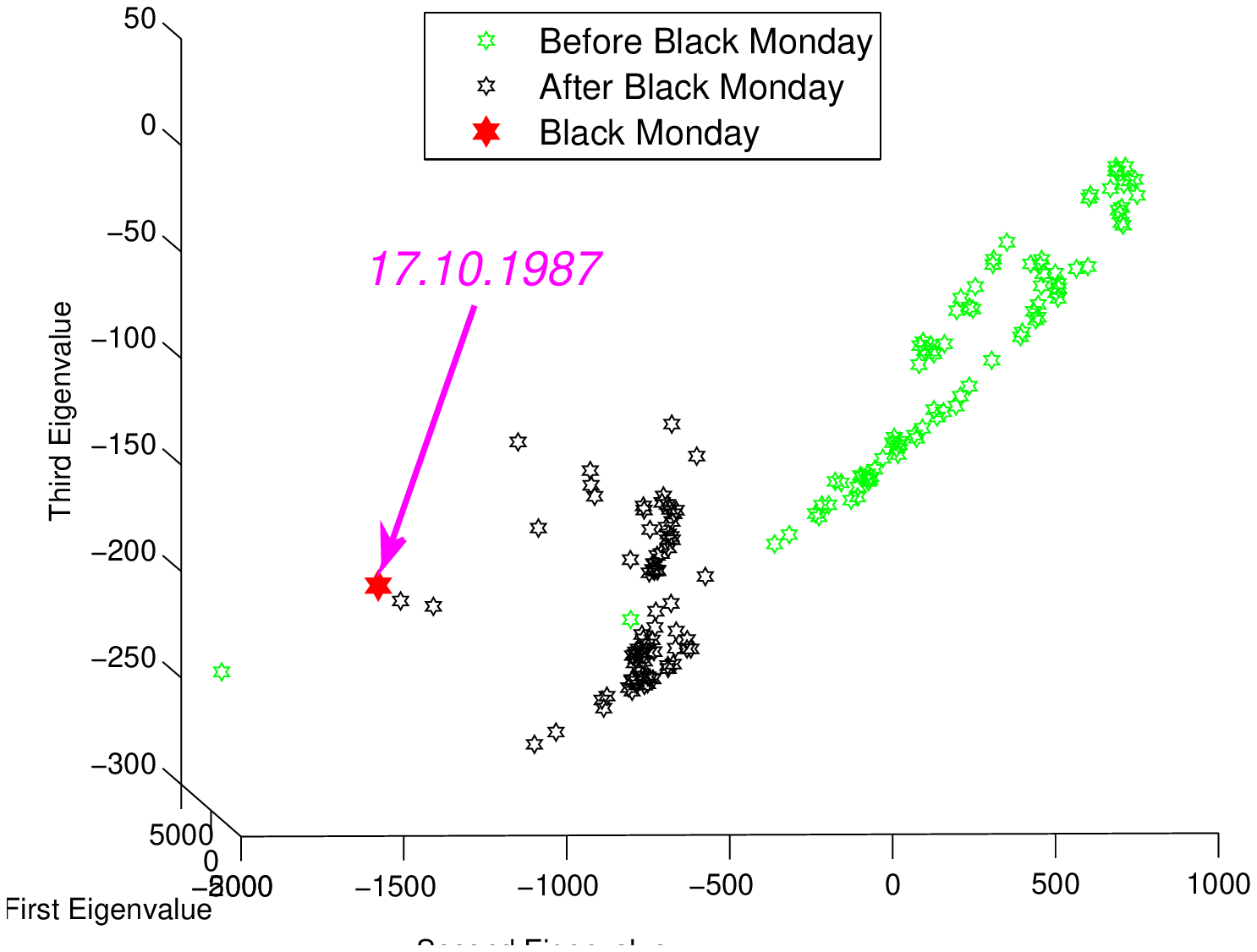}}
\subfigure[Dot-com Bubble]{\includegraphics[width=0.49\linewidth]{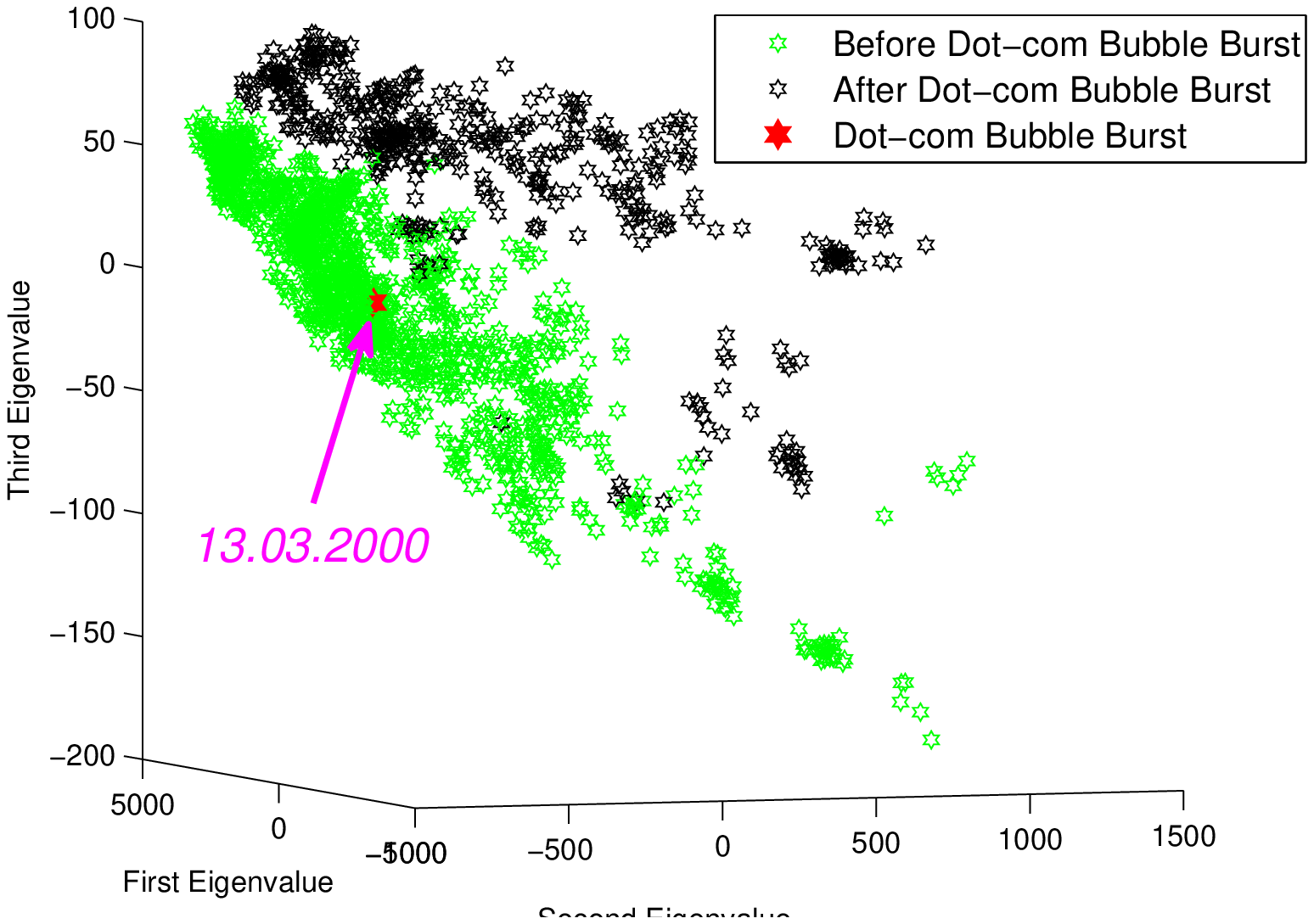}}
\subfigure[Enron Incident]{\includegraphics[width=0.49\linewidth]{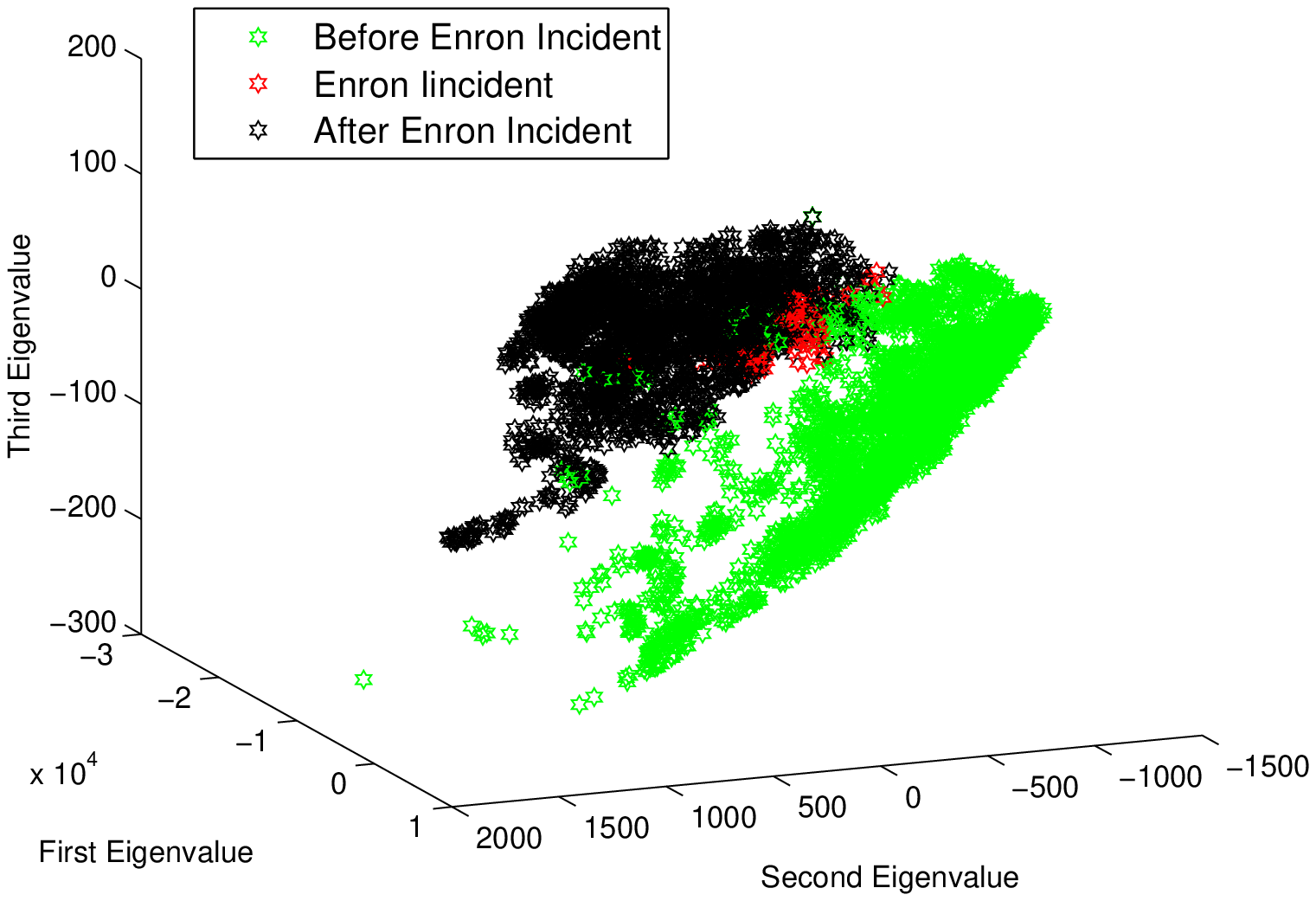}}
\subfigure[Subprime Crisis]{\includegraphics[width=0.49\linewidth]{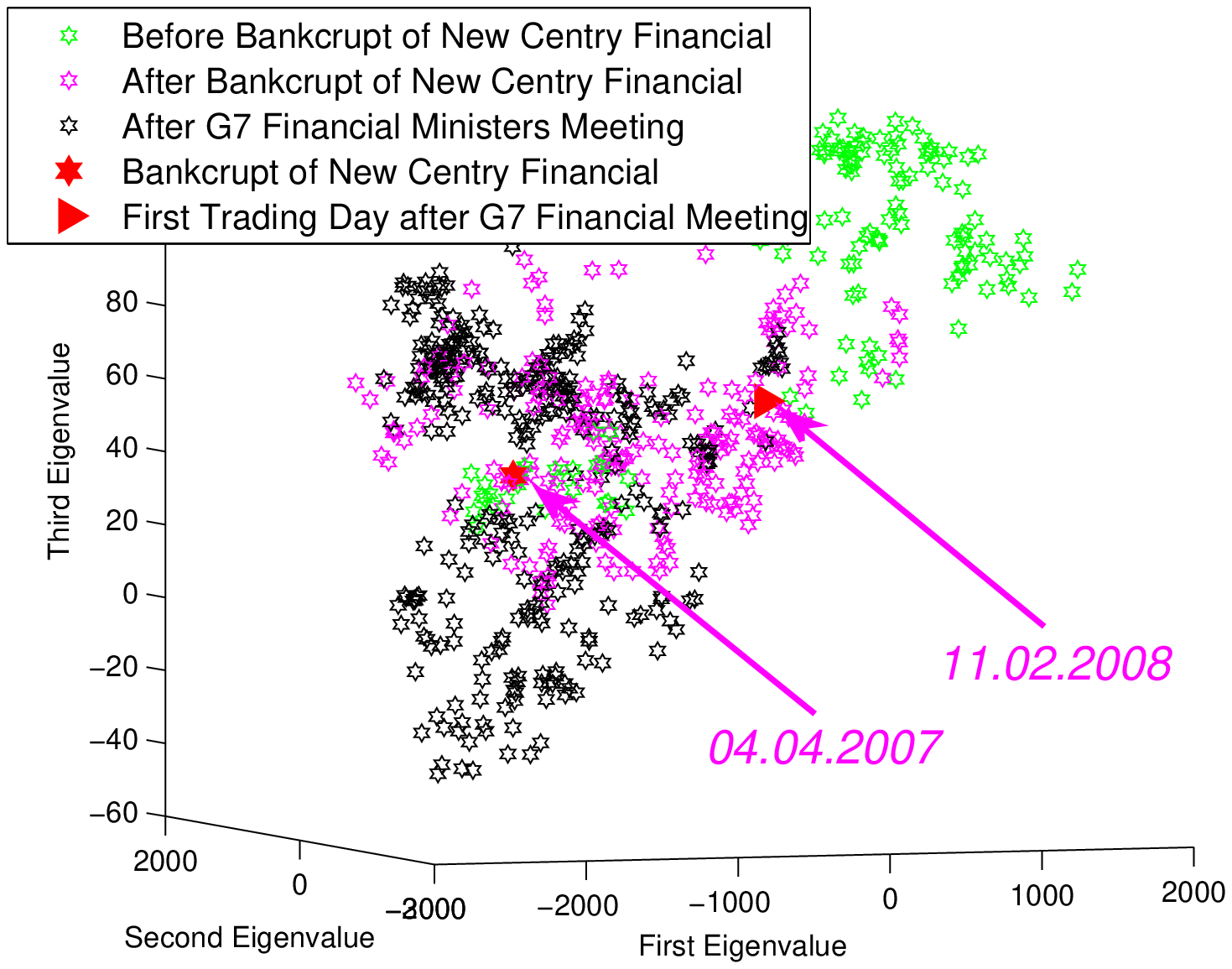}}
\vspace{-10pt}
\caption{The 3D embedding of the financial networks during different finance crises based on kPCA from JSGK (for original complete weight graphs).} \label{embeddingsCompareJS}
\vspace{-10pt}
\end{figure}

\begin{figure}
\centering
\subfigure[Black Monday]{\includegraphics[width=0.49\linewidth]{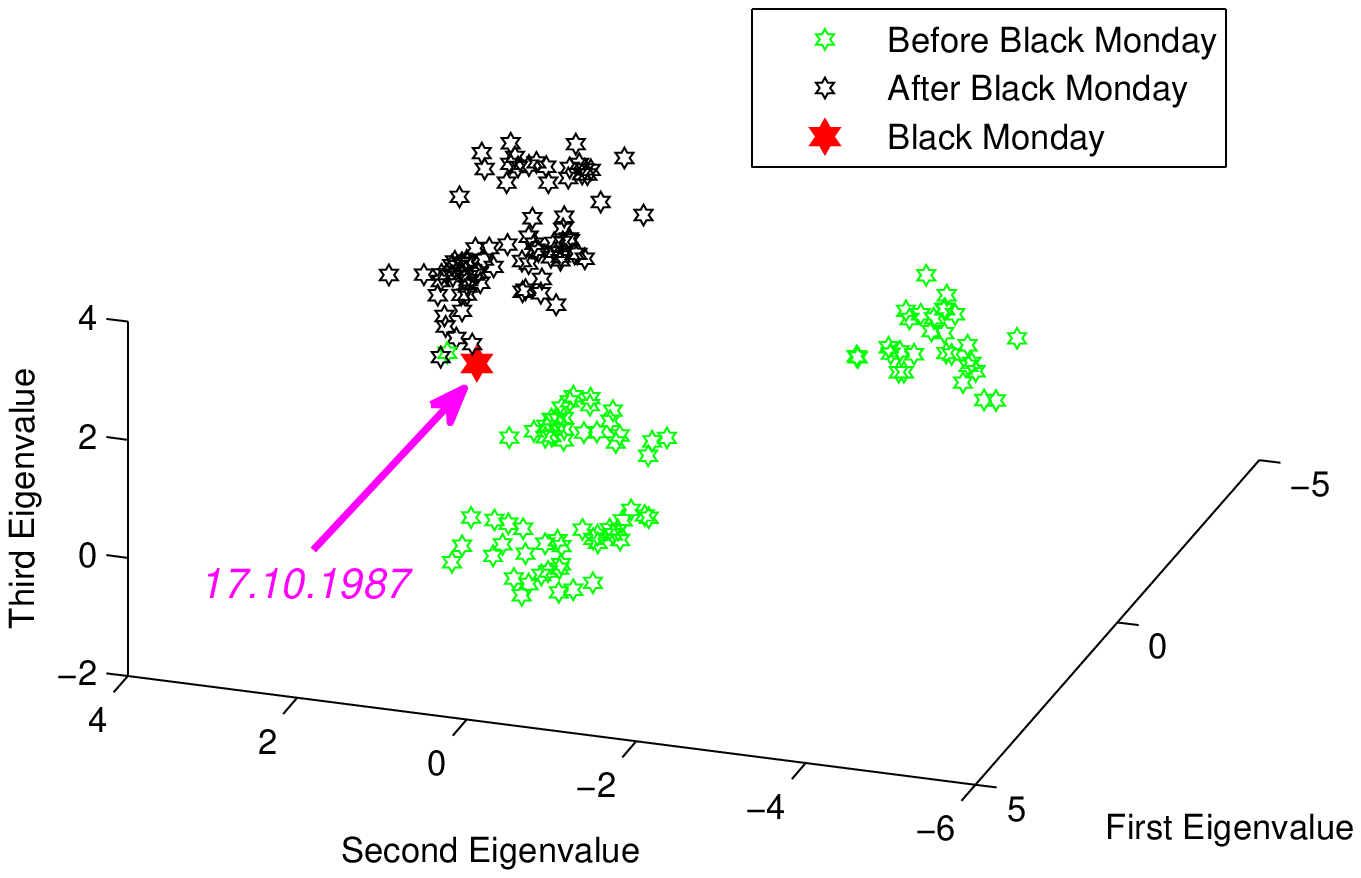}}
\subfigure[Dot-com Bubble]{\includegraphics[width=0.49\linewidth]{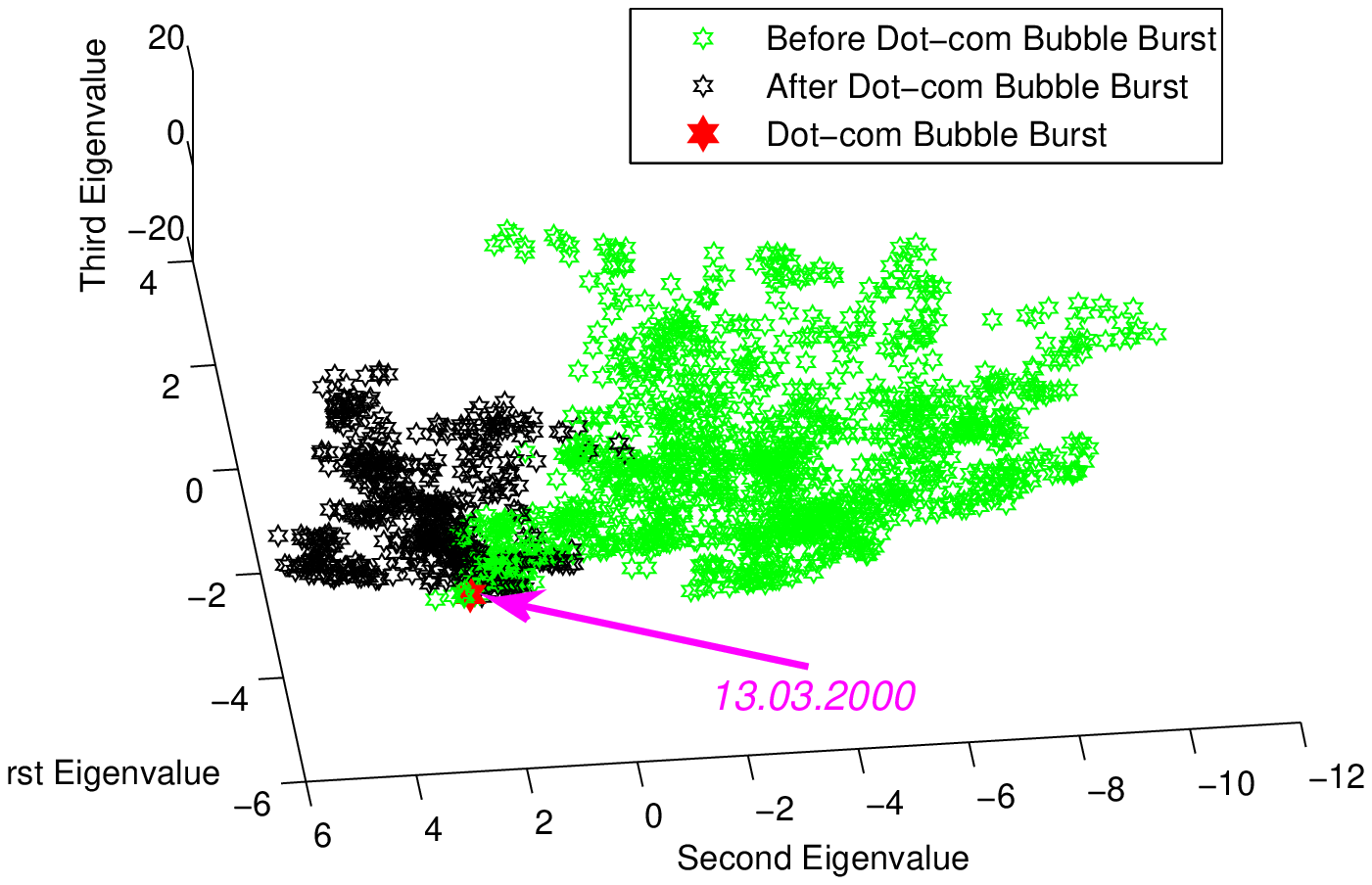}}
\subfigure[Enron Incident]{\includegraphics[width=0.49\linewidth]{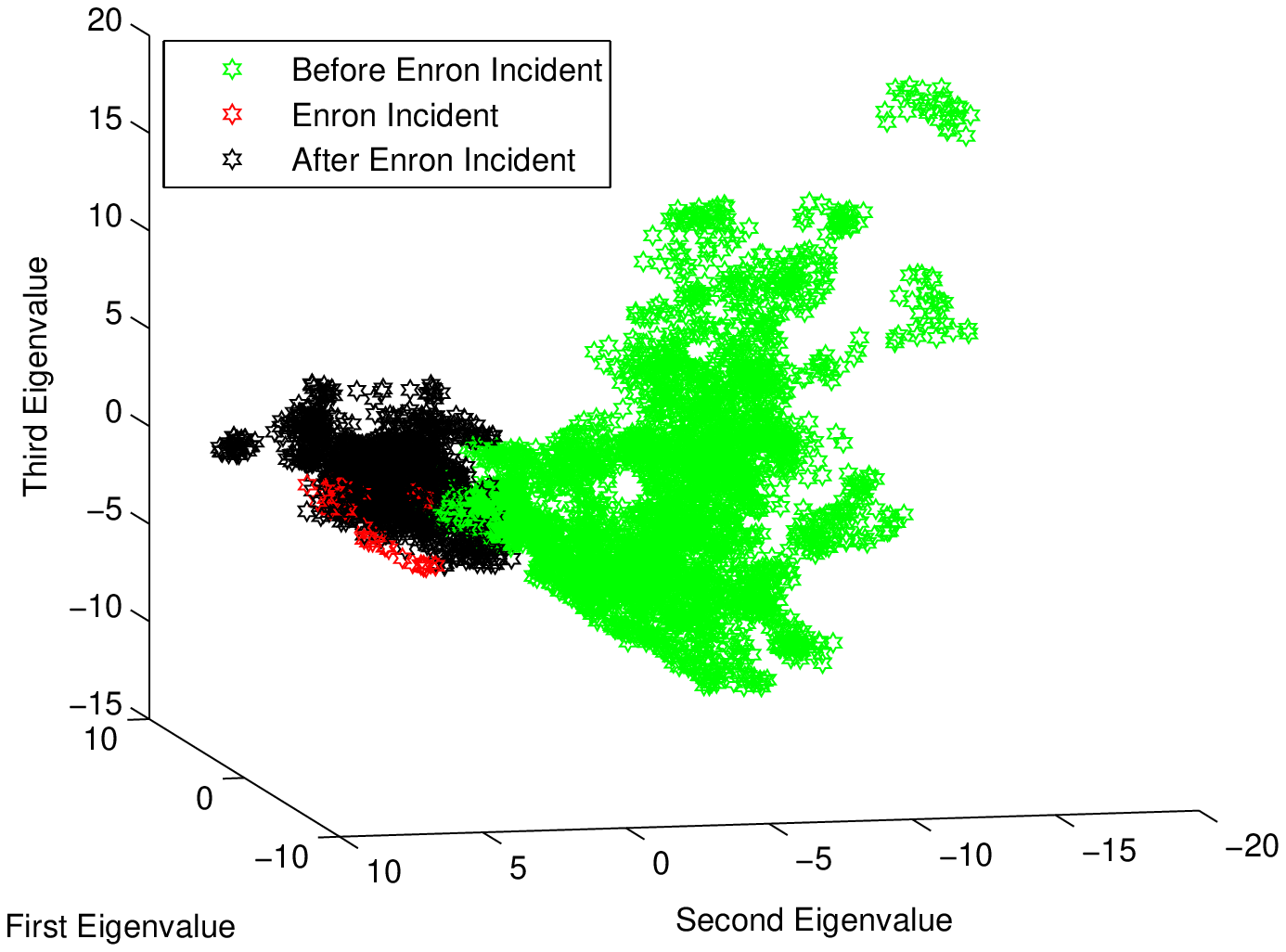}}
\subfigure[Subprime Crisis]{\includegraphics[width=0.49\linewidth]{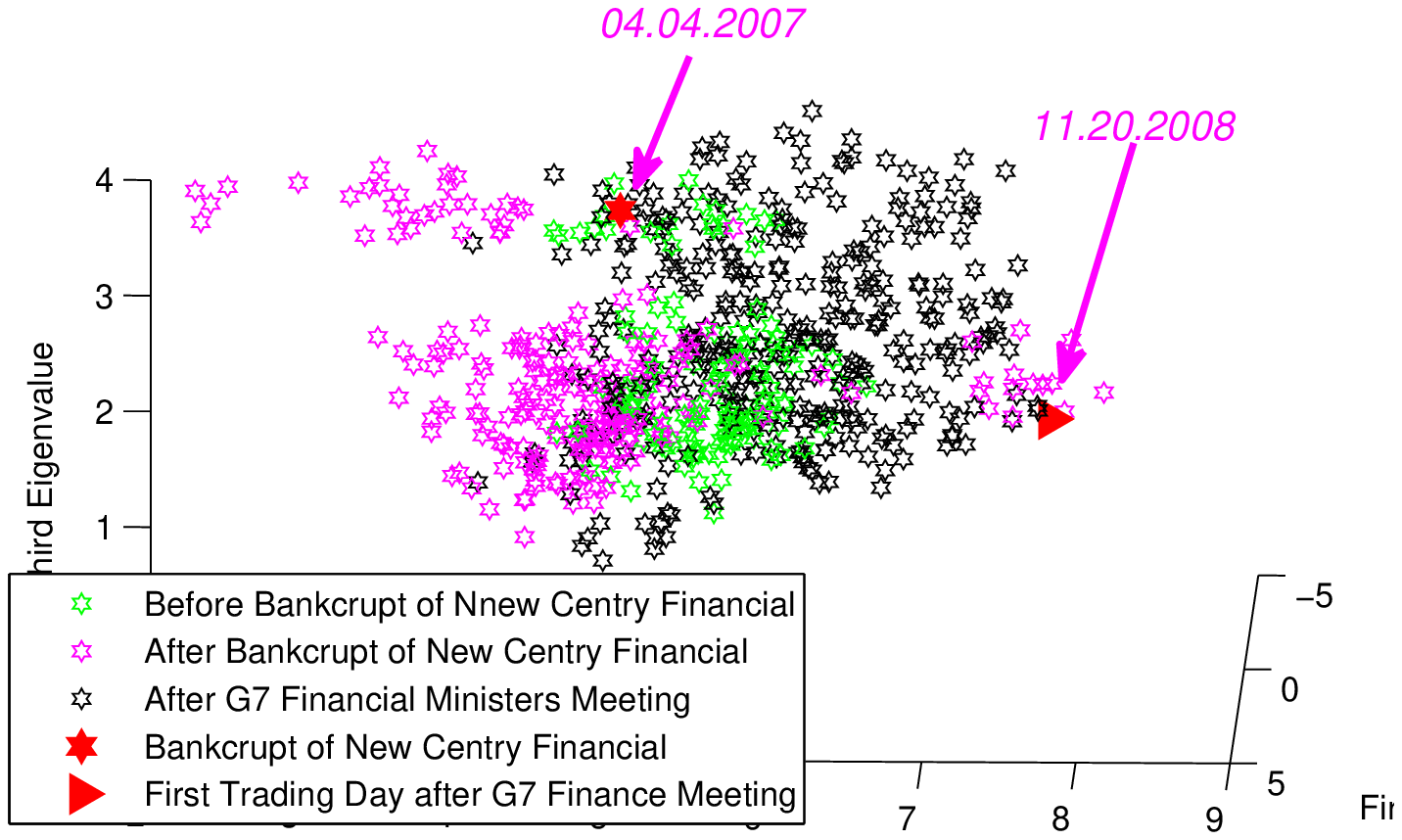}}
\vspace{-10pt}
\caption{The 3D embedding of the financial networks during different finance crises based on kPCA from QJS (for original complete weighted graphs).} \label{embeddingsCompareQJS}
\vspace{-10pt}
\end{figure}

\begin{figure}
\centering
\subfigure[Black Monday]{\includegraphics[width=0.49\linewidth]{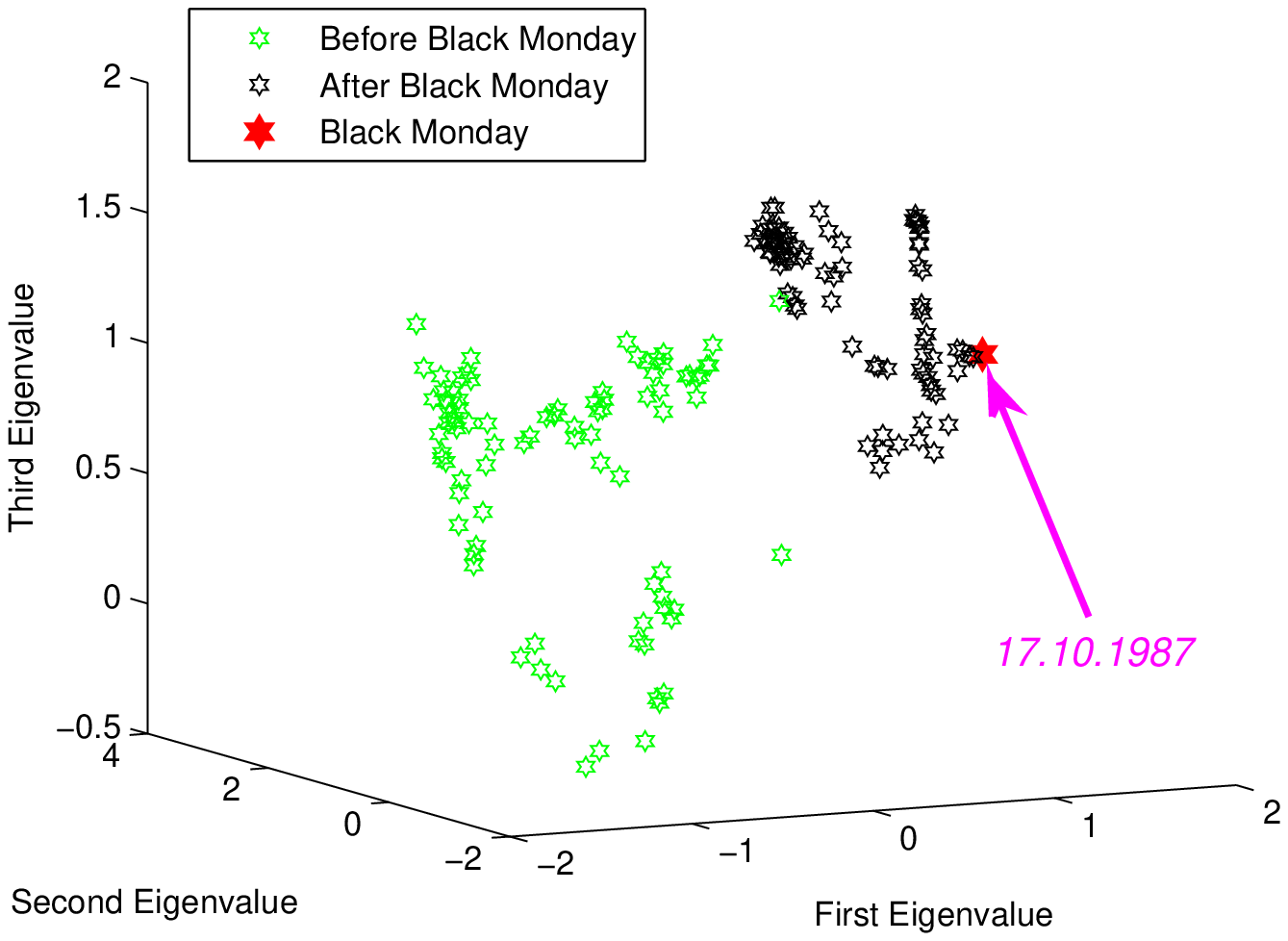}}
\subfigure[Dot-com Bubble]{\includegraphics[width=0.49\linewidth]{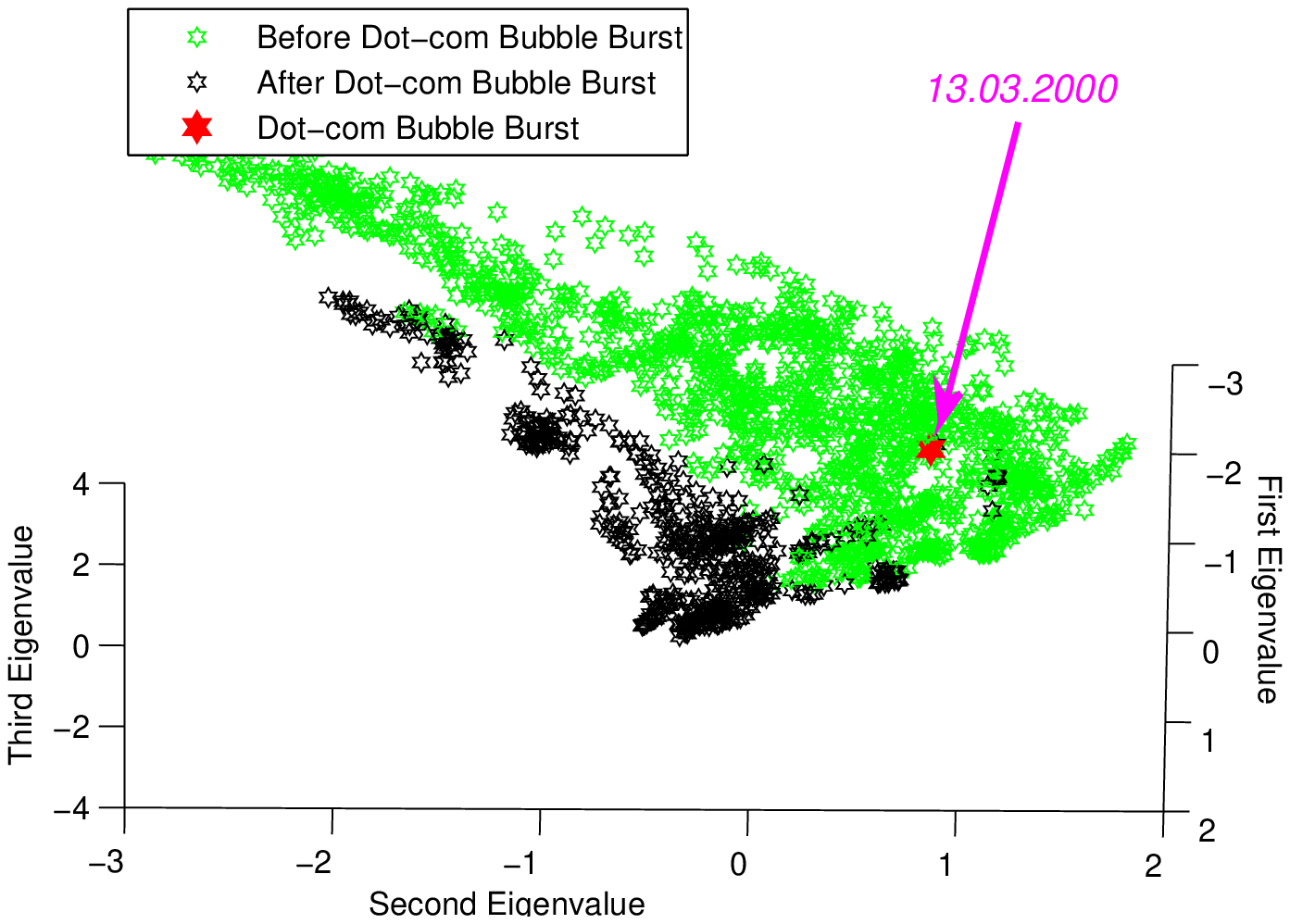}}
\subfigure[Enron Incident]{\includegraphics[width=0.49\linewidth]{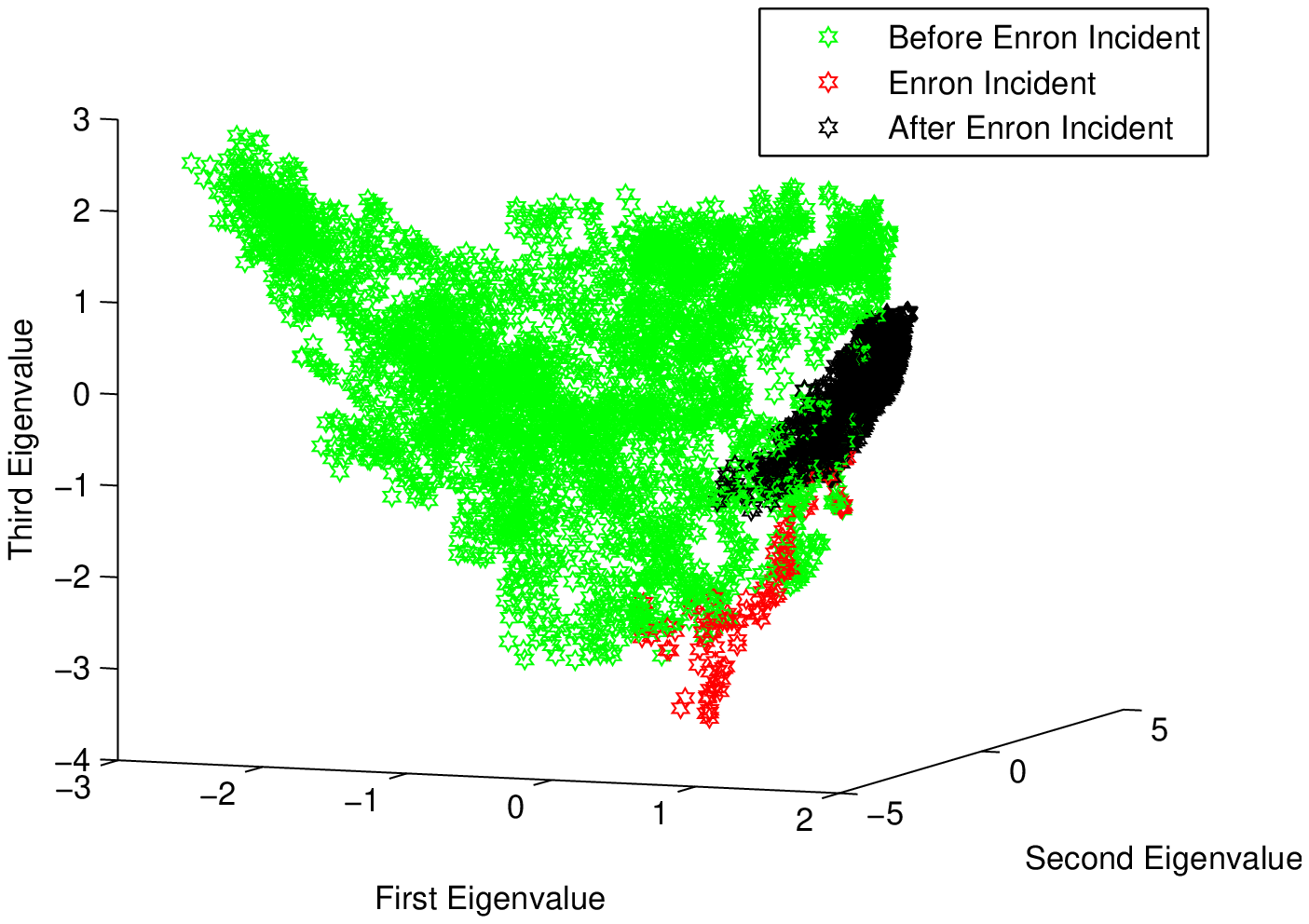}}
\subfigure[Subprime Crisis]{\includegraphics[width=0.49\linewidth]{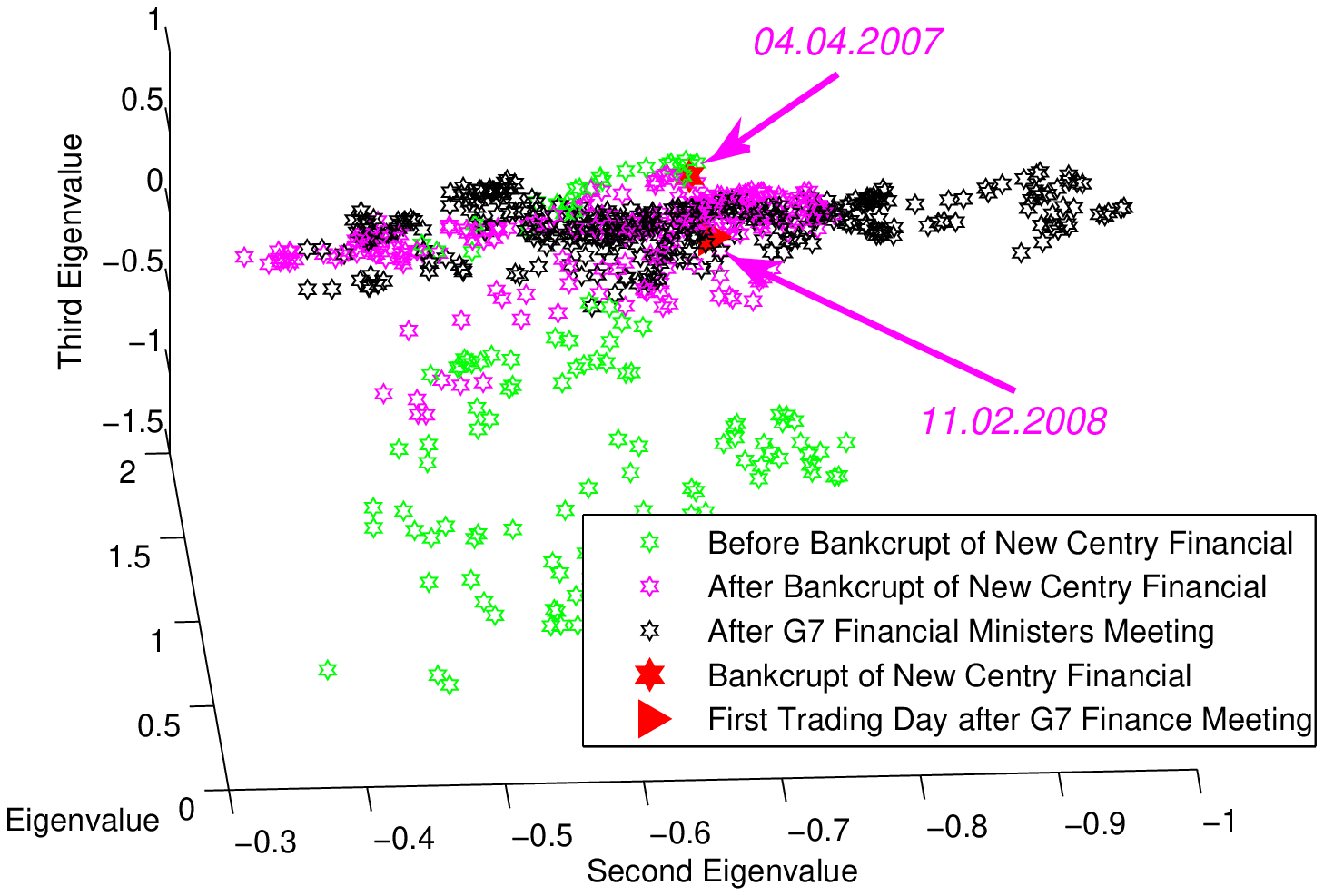}}
\vspace{-10pt}
\caption{The 3D embeddings of the financial networks during different crises based on kPCA from FLGK (for original complete weighted graphs).} \label{embeddingsCompareFLGK}
\vspace{-10pt}
\end{figure}

Similar embeddings are also found for a) the WLSK kernel in Fig. \ref{embeddingsCompareWL}, b) the JSGK kernel in Fig. \ref{embeddingsCompareJS}, c) the QJS kernel in Fig. \ref{embeddingsCompareQJS}, and d) the FLGK kernel in Fig. \ref{embeddingsCompareFLGK}. We observe that the proposed kernel outperforms the WLSK, JSGK, QJS and FLGK kernels in terms of the distributions of networks. More specifically, for the proposed quantum kernel, the boundaries between clusters are clear and the clusters are tighter. The reasons for this are threefold. First, unlike the proposed kernel, the WLSK kernel cannot encapsulate weight information residing on the edges. As a result, the WLSK kernel may lose important information from the financial networks. Second, the state space of the discrete-time quantum walk is larger than that of the steady state random walk and the continuous-time quantum walk, and thus can reflect more topological information. Moreover, the discrete-time quantum walk can limit the tottering problem arising in the classical random walk. As a result, the proposed quantum kernel can reflect richer characteristics than the JSGK and QJS kernels. Third, the discrete-time quantum walk can reflect more information of a network structure than its simple Laplacian matrix. As a result, the proposed kernel can better discriminate different financial network structures than the FLGK kernel. The above observations demonstrate that our kernel can distinguish different types of network evolution for time-varying financial networks, and outperforms state-of-the-art methods.

 % \subsection{Extra Financial Network Analysis in Supplementary Materials}
Since the commute time minimum spanning tree plays an important role in determining the performance of the proposed quantum kernel, we also provide experiments to demonstrate the advantage of using this kind of tree structures. Our experiments demonstrate that when the minimum spanning trees are abstracted through the commute time matrix, they can better preserve the graph characteristics than those abstracted through the original graph adjacency matrix. Moreover, to further demonstrate the effectiveness of the proposed kernel, we also evaluate its performance on time-varying financial networks where the edges are weighted according to the correlation instead of the Euclidean distance between the corresponding time series. The experimental results demonstrate again that the proposed kernel can outperform alternative methods on this kind of financial networks. Indeed, only the proposed kernel can simultaneously preserve the network characteristics in the Hilbert space and accommodate the edge weights. Due to the limit space of this manuscript, the results of these experimental evaluations are included in the supplementary materials.

\subsection{Experiments on General Graphs}
We conclude our experiments by showing the classification performance of the proposed kernels on more general standard graph datasets abstracted from the field of bioinformatics. These datasets include the MUTAG, PPIs(Proteobacteria40 PPIs and Acidobacteria46
PPIs), PTC(MR) and NCI1 datasets and their information is shown in Table~\ref{T:GraphInformation}. More introduction of these datasets can be found in~\cite{DBLP:journals/prl/Bai16}. Note that, unlike the time-varying networks, the graphs from these datasets are neither edge weighted nor completed graphs. To accommodate these graphs using the proposed kernels, we proceed as follows. For each graph, we first compute its commute time matrix, and assign each edge a weight using the commute time value between the pair of vertices connected by the edge, i.e., we transform each original graph into a weighted graph by assigning each original edge a weight using the corresponding commute time value. If the ratio between the edge and vertex number of the graph is larger than $1.5$ (i.e., the graph is dense), we compute the weighted minimum spanning tree over its weighted adjacency matrix as the new graph structure. We measure the kernel values using the proposed kernels between the original or sparsified structures. Furthermore we also strengthen the labels using the Weisfeiler-Lehman (WL) method~\cite{shervashidze2010weisfeiler} based on different iterations $h$. Recall that the WL method consists of the repeated propagation of the label information of a vertex to its neighbours, and when $h=0$ the strengthened vertex labels are the original vertex labels. In our experiments, we set the largest value of $h$ as 3, and compute the kernel matrix using each of the proposed kernels associated with all strengthened vertex labels by varying $h$ from 0 to 3. Note that, when no vertex labels are available, each vertex is first labelled with its degree before applying the WL strengthening approach. Finally, note that since each graph of the MUTAG dataset has an original discrete edge label, the required directed edge label is computed as in Eq.(\ref{edgelabelev}). For the remaining datasets, the required directed edge label is computed as in Eq.(\ref{edgelabele}).

\begin{table}[t!]
\vspace{0pt}
\centering {%
% \footnotesize
% \scriptsize
 \tiny
\caption{Summary statistics for the graph datasets.}\label{T:GraphInformation}
\vspace{-10pt}
\begin{tabular}{|c||c||c||c||c|}
  \hline
 ~Datasets ~         & ~MUTAG~  &  ~PPIs~    & ~PTC~    &~NCI1~ \\ \hline \hline
  ~Max \# vertices~  & ~$28$~   &  ~$232$~   &  ~$109$~ &~111~   \\
  ~Min \# vertices~  & ~$10$~   &  ~$3$~     &  ~$2$~   &~3~        \\
  ~Avg \# vertices~  & ~$17.93$~&  ~$109.60$~&  ~$25.60$~&~29.87~        \\  \hline

  ~Max \# edges~     & ~$33$~   &  ~$1503$~  &  ~$108$~  &~119~         \\
  ~Min \# edges~     & ~$10$~   &  ~$2$~     &  ~$1$~    &~2~          \\
  ~Avg \# edges~     & ~$19.79$~&  ~$432.18$~&  ~$25.96$~&~32.30~         \\  \hline

  ~\# graphs~        & ~$188$~  &  ~$86$~    &  ~$344$~  &~4110~         \\
  ~\# classes ~      & ~$2$~    &  ~$2$~     &  ~$2$~    &~2~         \\ \hline

 ~{Avg \# edges}/{Avg \# vertices}~& ~$1.10$~&~$3.94$~&~$1.01$~&~$1.08$~    \\ \hline

\end{tabular}
}
\vspace{-10pt}
\end{table}

\begin{table}
\centering {
\tiny
%\scriptsize
% \footnotesize
\caption{Accuracy (in $\%$ $\pm$ standard error) and runtime (in second).}\label{comparison}
\vspace{-10pt}
\begin{tabular}{|c||c||c||c||c|}
	
\hline
~Datasets~&~MUTAG~&~PPIs~&~PTC~&~NCI1~\\\hline
\hline

$\mathrm{K}_{\mathrm{DP}}$~&~$82.38\pm0.55$~&~$88.62\pm0.86$~&~$58.62\pm0.69$~&~$85.03\pm0.12$~\\\hline

$\mathrm{K}_{\mathrm{JS}}$~&~$\textbf{85.44}\pm0.58$~&~$\textbf{90.62}\pm0.90$~&~$\textbf{60.70}\pm0.62$~&~$\textbf{85.18}\pm0.10$~\\\hline

~WLSK~&~$82.05\pm0.57$~&~$78.50\pm1.40$~&~$56.05\pm0.51$~&~$80.68\pm0.27$~\\\hline

~QJSK~&~$83.83\pm0.49$~&~$70.57\pm1.20$~&~$58.23\pm0.80$~&~$67.40\pm0.20$~\\\hline

~QJSKT~&~$81.55\pm0.53$~&~$68.12\pm0.84$~&~$57.44\pm0.36$~&~$67.00\pm0.15$~\\\hline

~SPGK~&~$83.38\pm0.81$~&~$61.12\pm1.09$~&~$56.55\pm0.53$~&~$74.21\pm0.30$~\\\hline

~JSGK~&~$83.11\pm0.80$~&~$57.87\pm1.36$~&~$57.29\pm0.41$~&~$62.50\pm0.33$~\\\hline

~BRWK~&~$77.50\pm0.75$~&~$53.50\pm1.47$~&~$53.97\pm0.31$~&~$60.34\pm0.17$~\\\hline

\end{tabular}

\begin{tabular}{|c||c||c||c||c|}
\hline
~Datasets~&~MUTAG~&~PPIs~&~PTC~&~NCI1~\\\hline
\hline

$\mathrm{K}_{\mathrm{DP}}$~&~$2.7\cdot 10^1$~&~$5.5\cdot 10^1$~&~$2.7\cdot 10^1$~&~$5.4\cdot 10^2$~\\\hline

$\mathrm{K}_{\mathrm{JS}}$~&~$1.2\cdot 10^1$~&~$1.7\cdot 10^2$~&~$2.7\cdot 10^2$~&~$4.1\cdot 10^4$~\\\hline

~WLSK~&~$0.4\cdot 10^1$~&~$1.3\cdot 10^1$~&~$1.1\cdot 10^1$~&~$1.5\cdot 10^2$~\\\hline

~QJSK~&~$1.2 \cdot 10^1$~&~$1.4\cdot 10^4$~&~$1.1 \cdot 10^2$~&~$1.6 \cdot 10^4$~\\\hline

~QJSKT~&~$2.9\cdot 10^1$~&~$1.5\cdot 10^2$~& ~$1.7\cdot 10^2$~&~$1.4\cdot 10^4$~ \\\hline

~SPGK~&~$0.1\cdot 10^1$~&~$0.7\cdot 10^1$~&~$0.1\cdot 10^1$~&~$8.3\cdot 10^1$~\\\hline

~JSGK~&~$0.1\cdot 10^1$~&~$0.1\cdot 10^1$~&~$0.1\cdot 10^1$~&~$0.1\cdot 10^1$~\\\hline
~BRWK~&~$0.1\cdot 10^1$~&~$8.6\cdot 10^2$~&~$0.3\cdot 10^1$~&~$4.1\cdot 10^2$~\\\hline

\end{tabular}

}\vspace{-10pt}
\end{table}

We compare the performance of the proposed kernels $\mathrm{K}_{\mathrm{DP}}$ and $\mathrm{K}_{\mathrm{JS}}$ with that of several alternative state-of-the-art graph kernels. In addition to the kernels considered in the previous subsections, i.e., the WLSK~\cite{shervashidze2010weisfeiler}, the JSGK~\cite{DBLP:journals/jmiv/BaiH13}, the QJSK~\cite{DBLP:conf/gbrpr/Bai0RZH15}, and the QJSKT~\cite{DBLP:journals/prl/Bai16}, we consider the widely adopted shortest path graph kernel (SPGK)~\cite{DBLP:conf/icdm/BorgwardtK05}, and the backtrackless version of the random walk kernel (BRWK)~\cite{DBLP:journals/tnn/AzizWH13}. For each dataset, we report the average classification accuracies ($\pm$ standard error) and the time usage of computing the kernel matrices for each kernel in Table~\ref{comparison}. The results are computed by performing a $10$-fold cross-validation using a C-Support Vector Machine (C-SVM) to evaluate the classification accuracies of the different kernels. For each class, we used 90\% of the samples for training and the remaining 10\% for testing. The parameters of the C-SVMs are optimized separately on the training set for each dataset.

Table~\ref{comparison} shows that the proposed kernels outperform the alternative kernels on any dataset. Although our kernel measures are faster to compute than both the QJSK, the QJSKT, and the BRWK, the proposed kernel $\mathrm{K}_{\mathrm{JS}}$ through the Jensen-Shannon divergence has a significantly higher runtime than the WLSK kernel and the JSGK kernel. However our kernels can still complete the computation in polynomial time, while yielding a better classification performance. This effectiveness is due to the fact that only the proposed kernels can accommodate both the original edge and vertex labels, e.g, on the MUTAG dataset. Moreover, only the proposed kernels can accommodate the edge weights through the commute time and reflect more graph characteristics than the alternative kernels.

\section{Conclusion}\label{s6}
In this paper, we have proposed a new kernel measure for complete weighted graphs based on discrete-time quantum walks. Unlike existing state-of-the-art graph kernels, our kernel can accommodate complete weighted graphs, while at the same time overcoming the inefficiency and ineffectiveness of discrete-time quantum walks on these graphs. Experiments on time-varying financial networks abstracted from the New York Stock Exchange database as well as standard bionformatics graph datasets demonstrate the effectiveness of our kernel.

Our future work aims to extend the discrete-time quantum walk kernel for hypergraph-based time-varying financial networks. Ren et al.~\cite{DBLP:journals/pr/RenAWH11} have explored the use of the directed line graph representations for hypergraphs that can reflect richer high-order information than graphs. As we have stated, the discrete-time quantum walk can be seen as a walk evolving on directed line graphs. Thus, it would be interesting to extend these works by comparing the quantum walks on the directed line graphs associated with a pair of hypergraph-based time-varying financial networks.

% Finally, note that the graph structures abstracted from the NYSE dataset are graphs with unique vertex labels. On the other hand, Dickinson et al.,~\cite{DBLP:journals/paa/DickinsonBDK04} have explored the matching methods for the graphs with unique vertex labels. It would be interesting to employ the proposed kernels on more unique vertex label based graph structures in~\cite{DBLP:journals/paa/DickinsonBDK04}.

\section*{Acknowledgments}
This work is supported by the National Natural Science Foundation of China (Grant no. 61602535 and 61503422), the Open Project Program of the National Laboratory of Pattern Recognition (NLPR), and the program for innovation research in Central University of Finance and Economics. Corresponding author: Lixin Cui, E-mail: cuilixin@cufe.edu.cn.

%-------------------------------------------------------------------------

\balance

%-------------------------------------------------------------------------

\bibliographystyle{IEEEtran}
\bibliography{example_paper}

\end{document}